\documentclass[journal,12pt,onecolumn,draftclsnofoot]{IEEEtran}
\usepackage{graphicx,amsmath,float,amsthm,hyperref,subcaption,amssymb,color,mathdots,yhmath,listings,array}
\usepackage[utf8]{inputenc}
\usepackage[english]{babel}

\newtheorem{theorem}{Theorem}
\newtheorem{corollary}{Corollary}[theorem] 
\newtheorem{remark}{Remark}
\addto\captionsenglish{}
\newcommand\Tstrut{\rule{0pt}{2.6ex}}         % = `top' strut
\title{Real-Time CRLB based Antenna Selection in Planar Antenna Arrays}  
\author{Masoud Arash,~\IEEEmembership{Student member, IEEE}, Ivan Stupia,~\IEEEmembership{Member, IEEE}, Luc Vandendorpe,~\IEEEmembership{Fellow, IEEE}
\thanks{The authors are with the Institute of Information and Communication Technologies, Electronics and Applied Mathematics (ICTEAM), Universit\'e Catholique de Louvain, 1348 Louvain-la-Neuve, Belgium (E-mail: masoud.arash@uclouvain.be).}}
\begin{document}
	\maketitle
	\begin{abstract}
	Estimation of User Terminals' (UTs') Angle of Arrival (AoA) plays a significant role in the next generation of wireless systems. Due to high demands, energy efficiency concerns, and scarcity of available resources, it is pivotal how these resources are used. Installed antennas and their corresponding hardware at the Base Station (BS) are of these resources. In this paper, we address the problem of antenna selection to minimize Cramer-Rao Lower Bound (CRLB) of a planar antenna array when fewer antennas than total available antennas have to be used for a UT. First, the optimal antenna selection strategy to minimize the expected CRLB is proposed. Then, using this strategy as a preliminary step, we present a two-stage greedy antenna selection method whose goal is to minimize the instantaneous CRLB. The optimal start point of the greedy algorithm is presented alongside some methods to reduce the algorithm's computational complexity. Numerical results confirm the accuracy of proposed solutions. They demonstrate that the proposed antenna selection method only requires a small proportion of the total available antennas to accomplish a significant amount of the total performance, enhancing hardware utilization efficiency. Also, it is shown that the presented algorithm has a high error tolerance. 
	\end{abstract}
\section{Introduction}
Localization of UTs is an indispensable part of wireless systems. On the one hand, many applications, like auto-driving cars and health services, directly require continuous tracking of UT's locations. On the other hand, using location information, higher data rates with higher efficiencies can be transferred to the UTs \cite{wymeersch20175g}. The Angle of Arrival (AoA) information plays a key role in these applications.

A general criterion to evaluate the capability of a system to localize UTs is CRLB. Numerous works have studied the CRLB in different system models \cite{shahmansoori2017position,arash2021analysis}. Many parameters of a system affect the CRLB. In particular, the arrangement of antennas on the receiver side substantially influences the ability of AoA estimation. A linear \cite{Arash2020LE}, concentric \cite{li2019accurate}, or planar \cite{hu2014esprit} antenna array each has certain pros and cons. Planar antenna array provides wide aperture and uses assigned space for antenna array more efficiently (making them a suitable choice for Massive MIMO systems) \cite{adnan2017effects}. Also, studies for planar antenna arrays can be easily generalized to the newly emerging concept of Reconfigurable Intelligent Surfaces (RIS) \cite{bjornson2019massive,huang2018energy}.

Antenna selection is a solution for various challenges in multi-antenna systems. Energy efficiency concerns may limit the number of active antennas \cite{arash2017employing}, hardware shortage or complexity \cite{elbir2019joint}, or multi-casting services where different antennas provide different services to different UTs \cite{khan2019network}. It is shown that by using antenna selection, the system's energy efficiency can be improved up to $220\%$ \cite{Arash2020LE}. Also, in \cite{bhattacharya2021selection} it is shown that by using $20\%$ of total available antennas, more than $60\%$ of total localization accuracy can be obtained, resulting in improved hardware usage efficiency. Another motivation for antenna selection is that using all of the available antennas may surpass the required accuracy of the system \cite{godrich2011sensor}. In RIS and massive MIMO systems, antenna selection is used to balance performance and hardware complexity \cite{he2021reconfigurable,mendoncca2020antenna}. When fewer antennas than the total available antennas have to be used in a wireless network, active antennas can be selected to improve certain aspects of the system, e.g., the ability to estimate AoA with higher accuracy. 

In this regard, \cite{mulleti2020fast} proposes a neural network selection method to minimize the log-determinant of CRLB in a radar system with a linear array. The log determinant of the CRLB is used as the cost function, and the neural network is trained by feeding the actual AoA of training samples. In \cite{elbir2019cognitive} deep learning is used to minimize CRLB for cognitive radars with linear, circular, and randomly distributed antenna arrays. In \cite{bhattacharya2021selection} authors present a greedy algorithm for sensor selection in distributed radars where the objective function to minimize is the probability of error in a Maximum a Posteriori Localization (MAP) method. It is shown that a greedy algorithm can yield outstanding performance with low complexity and low run time. Another use of greedy algorithms to solve the combinatorial optimization problem of antenna selection is studied in \cite{godrich2011sensor} where the goal is to minimize minimum square error in a distributed system, using a knapsack problem formulation. In \cite{zhang2020antenna} authors introduced a real-time antenna selection strategy that uses a cyclic algorithm to minimize posterior CRLB (PCRLB) in a 2D scenario for a compact MIMO radar. The PCRLB is the optimization criterion as the prefiltering of the radar will not affect PCRLB. In \cite{wang2014adaptive} authors used a Dinkelbach type algorithm to minimize CRLB of AoA when desired parameters for estimation were AoA and azimuth of a UT. An unsupervised online learning method is used to minimize total CRLB for azimuth and elevation of a UT in \cite{aboutanios2021online}, with three orders of magnitude lower complexity compared to \cite{wang2014adaptive} to make it possible to run in real-time.

In works that aim to minimize CRLB, like \cite{wang2014adaptive,aboutanios2021online}, the actual AoAs are used in the optimization process, and this brings the question of how one can feed their proposed algorithm with AoAs when they are unknown. The main point in minimizing CRLB for AoA in the 3D scenario is that as it is a combined function of both antenna set and AoAs, without knowledge of AoAs, it will not be possible to minimize CRLB. If AoAs are known, there will be no need to minimize them. The primary contribution of this work is to address this issue to minimize the CRLB.

Greedy algorithms have been widely used for solving antenna selection problem \cite{mendoncca2020antenna,bhattacharya2021selection,godrich2011sensor}. A key feature of the greedy algorithm is its simplicity, making it possible to run it quickly. However, a major difference between employing a greedy algorithm in a compact planar antenna array with the distributed case, like the ones considered in \cite{bhattacharya2021selection,godrich2011sensor}, is that in the case of a compact array, the first three antennas have to be chosen together. At each step of the greedy algorithm, one antenna is selected. To do this, the $CRLB$ is checked for every possible antenna, and the one that results in the minimum $CRLB$ will be selected at the corresponding step. However, as the $CRLB$ is calculated for a planar array, the antenna set that is put in $CRLB$'s formula must compose a plane, otherwise, it will result in infinity. So, the minimum requirement to start the greedy algorithm is a set of antennas that create a plane. As the minimum number of points that create a plane is three, the start point of the greedy algorithm needs to be a set of three antennas that are not in a line. After that, the greedy algorithm can continue by selecting one antenna at each step without facing the infinite $CRLB$ problem.

In \cite{arash2021analysis}, we studied the problem of antenna selection for a planar array whose reference point is at its corner under Dense Multi-path Channel (DMC) model. A greedy-based antenna selection strategy whose cost function is the expected $CRLB$ was proposed with an optimal start point. With the help of the expectation (over all possible values of AoA), it is possible to remove the effects of the instantaneous AoA and minimize it. In the first part of this work, in continuation of our previous work \cite{arash2021analysis}, the $CRLB$ for a planar array whose reference point is at its center in the 3D setting is calculated where the effects of unknown elevation are also considered in the calculation. Then, (instead of using a greedy method), an optimal antenna selection strategy to minimize the expected CRLB of a planar antenna array is proposed. This antenna selection strategy only requires the possible range of AoA and its distribution to take the expectation from the CRLB. The advantage of this selection procedure is that the antenna selection priority can be determined prior to system implementation.

In the next part, to address the problem of lack of knowledge about AoA, we present a two-stage antenna selection procedure that aims to minimize the instantaneous CRLB of AoA. In the first step, a crude estimation of AoA is made using selected antennas to minimize expected CRLB. Then based on this estimation, a greedy algorithm with an optimal start point is employed to minimize the instantaneous CRLB. To reduce the computational complexity of the greedy algorithm, especially for its initial start set, we present the optimal start set for every value of AoA. Furthermore, by using the characteristics of the CRLB, we prove that in each step of the greedy algorithm, by searching only half of the available antennas, the contribution of all antennas can be evaluated. This proposition will further reduce the computational complexity of utilizing the greedy algorithm.

The contributions of the paper is summarized as the following: 
\begin{itemize}
	\item The optimal antenna selection strategy to minimize the expected CRLB of AoA estimation in planar antenna arrays in a 3D scenario is presented. 
	\item By using the antenna selection for minimizing the expected CRLB, a two-stage antenna selection method is developed that aims to minimize instantaneous CRLB of AoA.
	\item The presented algorithm is designed to have low complexity to be suited for real-time applications.To do this, the optimal start point of the greedy algorithm is analytically presented alongside a lemma to reduce the search pool.
\end{itemize}

\emph{Notation}: Boldface lower case is used for vectors, $\boldsymbol{x}$, and upper case for matrices, $\boldsymbol{X}$. $\boldsymbol{X}^H$ denotes conjugate transpose $\boldsymbol{X}$, $\mathbb{E}_x\{.\}$ is expectation operator w.r.t $x$, $j=\sqrt{-1}$, $\mid.\mid$ stands for absolute value of a given scalar variable, $\otimes$ is Kronecker product, $diag(\boldsymbol{x})$ is a diagonal matrix with the elements of vector $\boldsymbol{x}$ on the main diagonal, $\bigcup$ is union operator and $tr$ is trace operator. Also, $\boldsymbol{I}_K$ is $K\times K$ identity matrix.

The remaining of paper is organized as follows. In section \ref{systemmod} the system model is presented. In section \ref{CRLBsec} the CRLB of the presented system model is calculated. Section \ref{selection} presents both the antenna selection method to minimize expected CRLB and the proposed two phase antenna selection method that aims to minimize instantaneous CRLB. Finally, section \ref{NumericalR} studies the performance of the proposed methods through simulations.
\section{System Model} \label{systemmod}
\begin{figure}[t]
	\centering
	\includegraphics[width=0.6\textwidth]{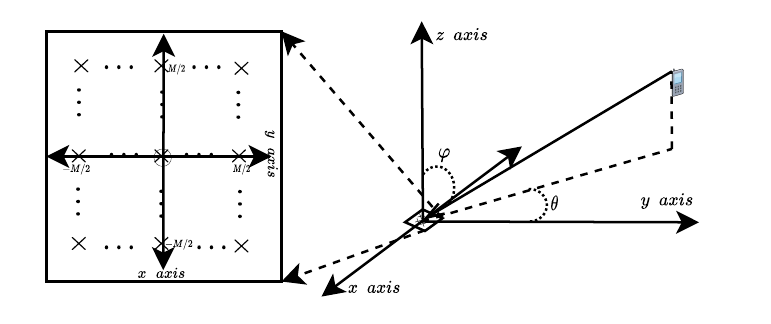}
	\caption{Antenna configuration at the BS w.r.t the UT.}
	\label{SysMod}
\end{figure}

We consider the uplink of a single-cell MIMO system with a BS at the center, equipped with $(M+1)^2$ antennas. $M+1$ antennas are installed along $x$ and $y$ axes. For mathematical tractability and clarity, it is assumed that $M$ is an even number. On each axis, adjacent antennas are separated by distance $d$ (Fig.\ref{SysMod}). There is a UT in the cell, and the BS uses pilot signals transmitted by the UT to localize the UT\footnote{Indeed, there can be more than one UT, but they have to use different frequency or time slot and the following concepts can be applied to each of them.}. For clarity, we assume that there is only one dominant path. The received signal at the BS is
\begin{equation}
	\boldsymbol{y}=\boldsymbol{a}_{R}h_ds+\boldsymbol{n},
\end{equation}
where $h_d$ is the channel coefficient for the dominant path of the UT, $s$ is the transmitted pilot by the UT with $\mathbb{E}\{s\}=0$ and $\mathbb{E}\{|s|^2\}=p_r$, and $\boldsymbol{n}\sim\mathcal{CN}(0,\sigma_n^2\boldsymbol{I}_M)$ is additive white Gaussian noise. Also, $\boldsymbol{a}_{Rx}$ is the steering vector of BS antenna array response, where $\theta$ and $\varphi$ are the UT's azimuth and elevation AoAs. From this point, azimuth AoA is simply called \emph{AoA} and elevation AoA is always referred to without any abbreviation. The $m$th element of $\boldsymbol{a}_{Rx}$ is \cite{hu2014esprit}
\begin{align}
	&\boldsymbol{a}_R(\theta,\varphi)_m=e^{-j\beta\sin(\varphi)((m_1-1)\cos(\theta)+(m_2-1)\sin(\theta))}, \label{StrP}\\
	&m=(m_2-1)(M+1)+m_1,\hspace{3mm} m_1=1,\ldots,M+1,\hspace{3mm} m_2=1,\ldots,M+1,\nonumber
\end{align}
where $\beta=\frac{2\pi d}{\lambda}$ and $\lambda$ is the wavelength of pilots.
\section{CRLB} \label{CRLBsec}
In this part, we obtain the formulation of $CRLB_\theta$ that is the CRLB for the estimation of the $\theta$, when $F\le(M+1)^2$ antennas are activated (and the rest of them are not used). As the CRLB of $\theta$ in a planar antenna array is well studied in the literature (namely in \cite{arash2021analysis,wang2014adaptive}), we just briefly explain how the same results can match our system model with minor changes in the $CRLB$ calculations\footnote{For detailed calculation, please refer to \cite{arash2021analysis}, section III.}. 

The vector of the desired variable is $\eta=[\theta\hspace{2mm}\varphi]$. Defining $\hat{\boldsymbol{\eta}}$ as the unbiased estimator of $\boldsymbol{\eta}$, its mean square error is lower bounded as 
\begin{equation}
	\mathbb{E}_{\boldsymbol{y}|\boldsymbol{\eta}}\{(\boldsymbol{\eta}-\boldsymbol{\hat{\eta}})(\boldsymbol{\eta}-\boldsymbol{\hat{\eta}})^T\}\ge \boldsymbol{CRLB}=\boldsymbol{J}^{-1}, \label{mse}
\end{equation}
where $\boldsymbol{J}$ is Fisher information matrix and can be written in block matrix form as \cite{shahmansoori2017position}
\begin{equation}
	\boldsymbol{J}=\left[\begin{array}{ccc} 
		J_{\theta,\theta}&J_{\theta,\varphi}\\
		J_{\varphi,\theta}&J_{\varphi,\varphi}
	\end{array}\right], \label{FIM}
\end{equation}
and 
\begin{equation}
	J_{a,b}=\mathcal{R}e[(\frac{\partial \boldsymbol{w}}{\partial \boldsymbol{\eta}_a})^H(\frac{\partial \boldsymbol{w}}{\partial \boldsymbol{\eta}_b})], \label{Jd}
\end{equation}
in which $\boldsymbol{w}\triangleq\boldsymbol{a}_{Rx}h_ds$ and $a,b\in\{\theta,\varphi\}$.
Therefore, using block matrix inversion properties, $CRLB_\theta$ is
\begin{equation}
	CRLB_\theta=\frac{\sigma_n^2}{2}(J_{\theta,\theta}-J_{\theta,\varphi}J_{\varphi,\varphi}^{-1}J_{\varphi,\theta})^{-1}. \label{crlb}
\end{equation}
From \cite{arash2021analysis}, equations (18-19), we know that
\begin{align}
	J_{\theta,\theta}=&\beta^2|h_{d}|^2\sin^2(\varphi)(tr(\tilde{\boldsymbol{\Sigma}}_1^2)\sin^2(\theta)+tr(\tilde{\boldsymbol{\Sigma}}_2^2)\cos^2(\theta)-2tr(\tilde{\boldsymbol{\Sigma}}_1\tilde{\boldsymbol{\Sigma}}_2)\cos(\theta)\sin(\theta)),\nonumber
\end{align}
\begin{align}
	J_{\varphi,\varphi}=&\beta^2|h_{d}|^2\cos^2(\varphi)(tr(\tilde{\boldsymbol{\Sigma}}_1^2)\cos^2(\theta)+tr(\tilde{\boldsymbol{\Sigma}}_2^2)\sin^2(\theta)+2tr(\tilde{\boldsymbol{\Sigma}}_1\tilde{\boldsymbol{\Sigma}}_2)\cos(\theta)\sin(\theta)), \nonumber
\end{align}
\begin{align}
	J_{\theta,\varphi}=&\beta^2|h_{d}|^2\sin(\varphi)\cos(\varphi)(tr(\tilde{\boldsymbol{\Sigma}}_2^2-\tilde{\boldsymbol{\Sigma}}_1^2)\sin(\theta)\cos(\theta)+tr(\tilde{\boldsymbol{\Sigma}}_1\tilde{\boldsymbol{\Sigma}}_2)(\cos^2(\theta)-\sin^2(\theta)). \label{Jtphi2}
\end{align}
where $\tilde{\boldsymbol{\Sigma}}_1$ and $\tilde{\boldsymbol{\Sigma}}_2$ are $F\times F$ selected matrices from $(M+1)^2\times (M+1)^2$  $\boldsymbol{\Sigma}_1$ and $\boldsymbol{\Sigma}_2$ matrices and 
\begin{align}
	&\boldsymbol{\Sigma}_1=\boldsymbol{I}_{M}\otimes diag(-M/2, \ldots, M/2), \nonumber\\
	&\boldsymbol{\Sigma}_2=diag(-M/2, \ldots, M/2)\otimes\boldsymbol{I}_{M}. 
\end{align}
The main points of obtaining \eqref{Jtphi2} from \cite{arash2021analysis} are two. First, due to absence of multi-path signals (caused by a different system model), the almost sure convergence in \cite{arash2021analysis} is replaced with equality in \eqref{Jtphi2}. Second, because there is only one UT, most of matrices are converted to scalars (including the $CRLB_\theta$). By defining $\rho\triangleq\frac{p_r}{\sigma_n^2}$ and replacing \eqref{Jtphi2} in \eqref{crlb} and after some algebraic operations, we obtain the $CRLB_{\theta}$ as
\begin{align}
	&CRLB_{\theta}=\frac{tr(\tilde{\boldsymbol{\Sigma}}_1^2)\cos^2(\theta)+tr(\tilde{\boldsymbol{\Sigma}}_2^2)sin^2(\theta)+2tr(\tilde{\boldsymbol{\Sigma}}_1\tilde{\boldsymbol{\Sigma}}_2)\cos(\theta)\sin(\theta)}{2\rho\beta^2|h_{d}|^2(tr(\tilde{\boldsymbol{\Sigma}}_1^2)tr(\tilde{\boldsymbol{\Sigma}}_2^2)-tr(\tilde{\boldsymbol{\Sigma}}_1\tilde{\boldsymbol{\Sigma}}_2)^2)\sin^2(\varphi)}. \label{CRLB}
\end{align}

By definition, $CRLB_\theta$ indicates the lowest possible error for any estimator for the considered system model. In this regard, $CRLB_\theta$ can be seen as an intrinsic characteristic of the system, as it is a function of system settings. Any unbiased estimator is bound to have a higher or equal variance as $CRLB_\theta$. This makes it a valuable parameter for various optimizations, namely antenna selection. If $CRLB_\theta$ is optimized, the performance of any estimator utilized in the system will be improved. Based on this, we choose $CRLB_\theta$ as the cost function of the antenna selection optimization problem, so its outcome can be used for any estimation method. 
\section{Antenna Selection} \label{selection}
In this section, the antenna selection in the localization phase is studied, and two antenna selection methods are proposed. In the first selection method, we use the expected value of $CRLB_\theta$ as the cost function to minimize. Then, using this method, we develop another antenna selection strategy whose cost function is instantaneous $CRLB_\theta$. We focus on the selection process and assume that the number of required antennas, $F$, is given.

Antenna selection is one of the proposed solutions in the literature to address hardware shortage \cite{romero2009performance}, or energy efficiency \cite{arash2017employing}. The general goal of antenna selection is to select a subset of antennas that results in the highest possible performance for the number of selected antennas. In \cite{arash2021analysis} we have proposed a greedy antenna selection strategy that aims to minimize expected $CRLB_\theta$ for a planar antenna array, where the expectation is over all possible realizations of AoA. 

In \cite{wang2014adaptive} authors presented an antenna selection to minimize $CRLB_\theta$ for a planar antenna array whose reference point is at the center of the array. The antenna set is selected based on the actual AoAs using the Dinkelbach algorithm for minimization. The problem with this approach is that AoAs have to be known before minimization. This means that the antenna selection cannot be used for the AoA estimation phase. Moreover, the complexity of the Dinkelbach algorithm may increase the time consumption of the antenna selection procedure. To overcome these problems, we propose a real-time antenna selection approach.

Our method is composed of two phases. The first phase is when the transmission of pilot signals by UT has not yet begun. In this phase, the only available information at the BS is the possible range and distribution of the AoA. For example, if the space division multiplexing is not used and the BS has to cover all of the cell, this range would be $[0, 2\pi)$. In this step, we turn on $F_p$ antennas to make a crude estimation of the received AoA to have a general idea about where the UT is. The value of $F_p$ can be defined by other criteria, like the total energy consumption of the BS. These $F_p$ antennas are selected to minimize the expected $CRLB_\theta$ over the possible range of the AoA. In this step, we will use a certain percentage of the total transmitted pilot snapshots to acquire a preliminary estimation of the AoA (we call this estimation $\theta_{p}$). 

Minimizing \eqref{CRLB} is a Combinatorial optimization problem. After acquiring the $\theta_{p}$, we can use it to minimize the $CRLB_\theta$. By virtue of the $\theta_{p}$, we can select a subset of the antennas that specifically minimize the $CRLB_\theta$ instead of its expectation. We use a greedy algorithm and present its optimal starting point. In the greedy approach, one antenna that reduces the $CRLB_\theta$ more than others is selected in each step. The greedy algorithm is chosen for the optimization due to two main reasons. First, it can be run in real-time and very fast, thanks to its simplicity. The running time is significant because this part of the selection process is done while BS receives the pilot signals. Extra delay in the selection process will result in missing transmitted snapshots by the UT. Second, the greedy algorithm has a high level of performance. Later in the numerical result section, we show that the greedy algorithm (with the optimal starting point) obtains almost the same performance as the global search with much lower computational complexity. 

Therefore, the steps of the antenna selection algorithm can be summarized as the following: 
\begin{enumerate}
	\item \textit{Preliminary stage}: $F_p\ge3$ antennas are selected in a way to minimize $\mathbb{E}_\theta\{CRLB_\theta\}$. BS turns on these antennas and awaits the reception of the pilot signals from the UT. Then, it uses these antennas alongside a percentage of the total snapshots to make a Preliminary estimation of $\theta$ (that is called $\theta_p$).
	\item \textit{Main stage}: Based on the $\theta_p$, a set of antennas are selected to minimize $CRLB_{\theta}$ using a greedy algorithm. This set will use the remaining snapshots of the pilot signals to estimate $\theta$.  
\end{enumerate}

\begin{remark}
	In addition to helping the antenna selection procedure, $\theta_p$ can be used to reduce the computational complexity of the estimation of $\theta$ in the Main stage. For example, in methods like ML or MUSIC, where a certain range has to be searched for the estimation, only the vicinity of $\theta_p$ (with a certain margin) can be the prime candidate to be searched. This reduces the search area and reduces estimation time. 
\end{remark}

In the following subsections, the antenna selection of each step is explained.
\subsection{Preliminary Stage} \label{PS1}
In this stage, the only available information is the total possible range of $\theta$. As $CRLB_\theta$ is a function of both antennas set and $\theta$, we have to remove its dependency to $\theta$ (based on the available information) to be able to use $CRLB_\theta$ as a cost function for minimization. In order to do so, by noting that $\theta\sim U[0,2\pi)$, the expectation of $CRLB_\theta$ over $\theta$ will be

\begin{align}
	\mathbb{E}_\theta\{CRLB_\theta\}=\frac{1}{2\rho\beta^2|h_{d}|^2\sin^2(\varphi)}\times \underbrace{\left(\frac{tr(\tilde{\boldsymbol{\Sigma}}_{p_1}^2)+tr(\tilde{\boldsymbol{\Sigma}}_{p_2}^2)}{(tr(\tilde{\boldsymbol{\Sigma}}_{p_1}^2)tr(\tilde{\boldsymbol{\Sigma}}_{p_2}^2)-tr(\tilde{\boldsymbol{\Sigma}}_{p_1}\tilde{\boldsymbol{\Sigma}}_{p_2})^2)}\right)}_{U(\mathcal{S})}, \label{ECRLB}
\end{align}
where $\tilde{\boldsymbol{\Sigma}}_{p_1}$ and $\tilde{\boldsymbol{\Sigma}}_{p_2}$ contain the corresponding values of $F_p$ selected antennas from $\boldsymbol{\Sigma}_{1}$ and $\boldsymbol{\Sigma}_{2}$, respectively and $\mathcal{S}$ is the set of selected antennas. The only part related to the antenna set in \eqref{ECRLB} is $U(\mathcal{S})$, so we have to minimize it. First, we explain the process in detail, and then we present and prove it in a theorem. 

\begin{figure}[t]
	\centering
	\includegraphics[width=0.5\columnwidth]{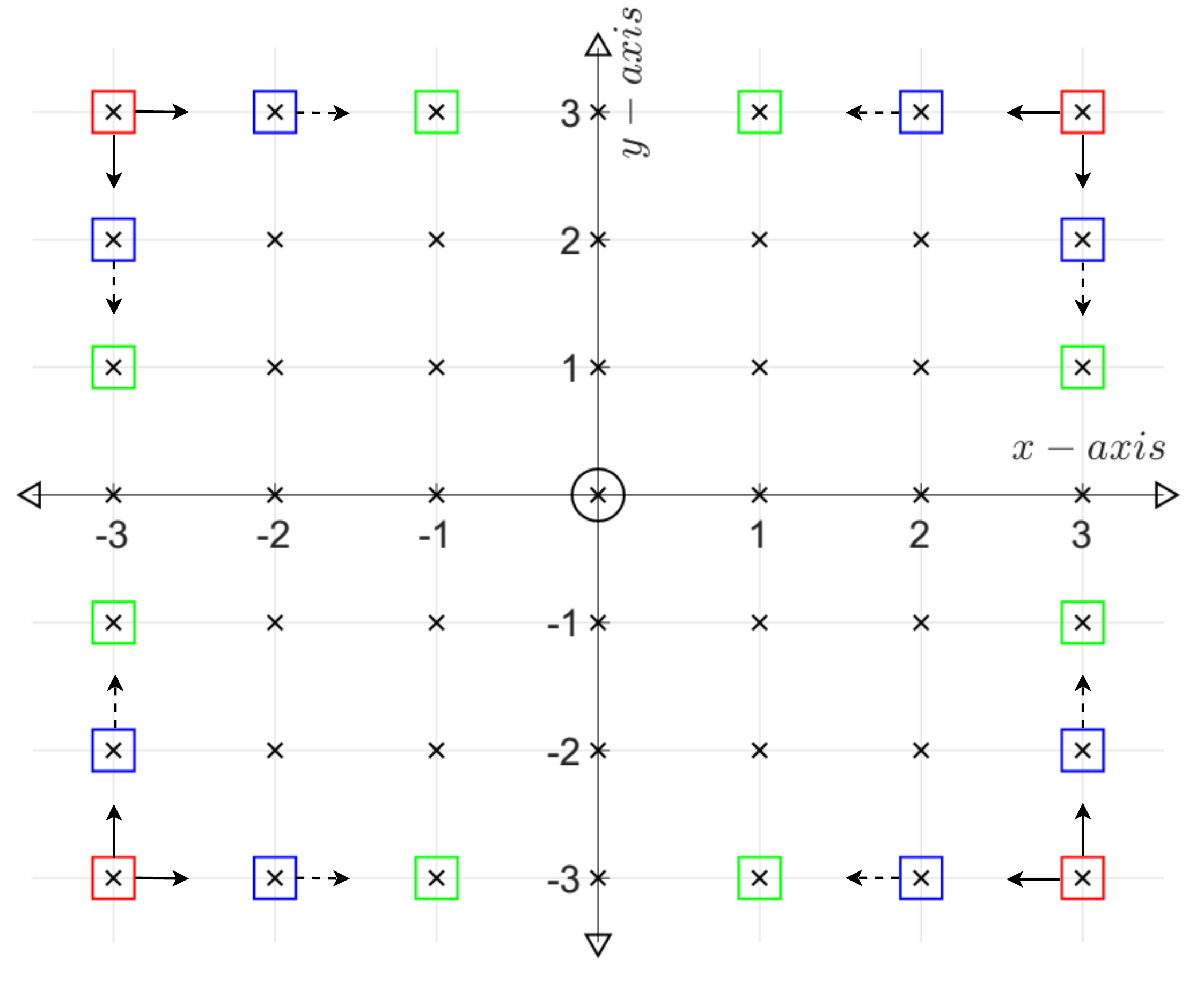}
	\caption{Selection Process for $\theta_{p}$ estimation stage.}
	\label{SelectionProc}
\end{figure}

The optimal antenna selection strategy to minimize $U(\mathcal{S})$ is composed of various steps that depend on the value of the $F_p$. If $F_p\le5$, in addition to the reference point, the four antennas at the corners of the array (i.e. $S_{p_1}=\{(\frac{M}{2},\frac{M}{2}), (-\frac{M}{2},\frac{M}{2}), (\frac{M}{2},-\frac{M}{2}), (-\frac{M}{2},-\frac{M}{2})\}$) are optimal choices and all of them have the same priority to be selected. This means that if, for example, $F_p=4$, either combination of three antennas from $S_{p_1}$ (in addition to the reference point) will result in the same value of $U(\mathcal{S})$. So, for $F_p=4$ any of the following sets are optimal choices with the same values for $U(\mathcal{S})$:
\begin{align}
	 &\{(0,0),(\frac{M}{2},\frac{M}{2}),(-\frac{M}{2},\frac{M}{2}),(\frac{M}{2},-\frac{M}{2})\},\nonumber\\ &\{(0,0),(-\frac{M}{2},-\frac{M}{2}),(-\frac{M}{2},\frac{M}{2}),(\frac{M}{2},-\frac{M}{2})\},\nonumber\\ &\{(0,0),(\frac{M}{2},\frac{M}{2}),(-\frac{M}{2},-\frac{M}{2}),(\frac{M}{2},-\frac{M}{2})\},\nonumber\\ &\{(0,0),(\frac{M}{2},\frac{M}{2}),(-\frac{M}{2},\frac{M}{2}),(-\frac{M}{2},-\frac{M}{2})\}\nonumber.
 \end{align}
For $5<F_p\le13$, the next optimal choices are neighboring antennas of the four antennas at $S_{p_1}$ in the outer most rectangle of the array, i.e. $S_{p_2}=\{(\frac{M}{2}-1,\frac{M}{2}), (1-\frac{M}{2},\frac{M}{2}), (\frac{M}{2},\frac{M}{2}-1), (-\frac{M}{2},\frac{M}{2}-1), (\frac{M}{2},1-\frac{M}{2}), (-\frac{M}{2},1-\frac{M}{2}),(\frac{M}{2}-1,-\frac{M}{2}), (1-\frac{M}{2},-\frac{M}{2})\}$. Again, all of the antennas at $S_{p_2}$ have the same priority to be selected, and if $5<F_p<13$, any combination from $S_{p_2}$ (in addition to all of the antennas in $S_{p_1}$ and the reference point) will result in the same value for $U(\mathcal{S})$. For $13<F_p\le21$, the neighbors of the ones selected for $5<F_p\le13$ in the outermost rectangle of the array are optimal choices and so on. Therefore, the process is to start with the four antennas at $S_{p_1}$ at the corners, select their neighbor antennas, and move toward other corners. An example of this process for $M=6$ is represented graphically in Fig.~\ref{SelectionProc}. In this figure, four red antennas have the highest priority to be selected. After selecting them, we move in both available directions (shown by solid arrows) to select antennas with blue markers. All antennas with blue markers have the same priority. Then the same procedure is continued by moving in the same directions (showed by dashed arrows) to obtain the next set of antennas, marked by green markers. When all antennas from the outermost rectangle are selected, the same process is repeated from the second outermost rectangle. The following theorem summarizes the optimal selection procedure. 

\begin{theorem} \label{T1}
	The optimal antenna selection strategy to minimize $\mathbb{E}_\theta\{CRLB_\theta\}$ for $\theta\sim U[0,2\pi)$ is to start from the four antennas at the corner of the array and then, from each corner, move toward other two corners, selecting antennas from the outermost rectangle. Antennas that have the same distance from their closest corner antenna, have the same priority for selection. When all antennas at the outermost rectangle are selected, the same procedure is repeated in the second outermost rectangle and so on. 	
\end{theorem}
\begin{proof}
	Please see Appendix \ref{App1}.
\end{proof}

Theorem\ref{T1} presents the optimal selection strategy when the goal is to minimize $\mathbb{E}_\theta\{CRLB_\theta\}$. This is an extension of our previous work in \cite{arash2021analysis} where we used a greedy algorithm to minimize $\mathbb{E}_\theta\{CRLB_\theta\}$. This strategy itself can be used as an independent selection method. The most important advantage of this method is that the desired set can be determined offline, and there is no need for online calculations, which leads to fast and easy implementation. For the rest of the paper, this method of antenna selection is referred to as \emph{expected} method.  

Another important aspect of Theorem\ref{T1} is the shape of the selected antennas. When $F\le4(M-1)$, the optimal choice creates four subarrays of antennas. This is in accordance with the same results presented in \cite{hu2018beyond} for large intelligent surfaces, which shows that four smaller subarrays perform better than one big antenna array for AoA estimation. We saw the same results in \cite{arash2021analysis} when the greedy algorithm is utilized to select antennas, where the goal is to minimize the $\mathbb{E}_\theta\{CRLB_\theta\}$. Furthermore, we will see in Section\ref{NumericalR} that for the real-time antenna selection, where the goal is to minimize $CRLB_\theta$, the greedy algorithm will again create smaller subarrays that are separated from one another. 
\subsection{Main Antenna selection}
In this part, we select the total number of desired antennas based on the $\theta_{p}$, acquired in the previous step. We use a greedy algorithm in this stage to minimize the $CRLB_\theta$. Because $CRLB_\theta$ in \eqref{CRLB} is for a planar antenna array, the greedy algorithm requires a starting point composed of three antennas selected simultaneously. This means that the first three antennas (the minimum number to form a plane) must be selected together and cannot be selected in a greedy approach. The search for selecting these three antennas together to initiate the greedy algorithm is usually very costly in terms of computational complexity. This is because one has to go through all possible combinations of choosing three antennas over $(M+1)^2$. However, with the help of the following theorem, this complexity will be reduced to only two possible choices for any $M$. In the following Theorem, we provide the optimal starting set for the greedy algorithm and prove that this set is the best possible choice. The coordinates of the optimal choices are extracted from the pattern observed in the simulations.
\begin{theorem} \label{T2}
	The optimal first three antennas that minimize $CRLB_\theta$ for different values of $\theta$ in the first quarter of the trigonometric circle are as the following:\\	
	If $0\le\theta\le\frac{\pi}{4}-\theta_0$,
	\begin{align}
		\mathcal{S}_3=\{(0,0),(-\frac{M}{2}, \frac{M}{2}),(x_2, \frac{M}{2})\} \label{1regopt},
	\end{align} 
	where
	\begin{equation*}
		x_2 = \left\{
		\begin{array}{rl}
			\lfloor \frac{M(1-2\alpha)}{2}\rfloor & \text{if } g_x(\lfloor \frac{M(1-2\alpha)}{2}\rfloor) \le g_x(\lceil \frac{M(1-2\alpha)}{2}\rceil)\\
			\lceil \frac{M(1-2\alpha)}{2}\rceil & \text{if } g_x(\lfloor \frac{M(1-2\alpha)}{2}\rfloor) > g_x(\lceil \frac{M(1-2\alpha)}{2}\rceil),\\
		\end{array} \right.
	\end{equation*}
	If $\frac{\pi}{4}-\theta_0<\theta<\frac{\pi}{4}+\theta_1$,
	\begin{align}
		\mathcal{S}_3=\{(0,0),(-\frac{M}{2}, \frac{M}{2}-1),(1-\frac{M}{2}, \frac{M}{2})\}, \label{pi/4}
	\end{align} 
	If $\frac{\pi}{4}+\theta_1\le\theta\le\frac{\pi}{2}$,
	\begin{align}
		\mathcal{S}_3=\{(0,0),(-\frac{M}{2}, \frac{M}{2}),(-\frac{M}{2}, y_2)\},
	\end{align} 
	where
	\begin{equation*}
		y_2 = \left\{
		\begin{array}{rl}
			\lfloor \frac{M(2-\alpha)}{2\alpha}\rfloor & \text{if } g_y(\lfloor \frac{M(2-\alpha)}{2\alpha}\rfloor) \le g_y(\lceil \frac{M(2-\alpha)}{2\alpha}\rceil)\\
			\lceil \frac{M(2-\alpha)}{2\alpha}\rceil & \text{if } g_y(\lfloor \frac{M(2-\alpha)}{2\alpha}\rfloor) > g_y(\lceil \frac{M(2-\alpha)}{2\alpha}\rceil),\\
		\end{array} \right.
	\end{equation*}
	$\alpha=\tan(\theta)$, 
	\begin{align}
		&g_x(x)=\frac{(1-\alpha)^2+(\frac{2}{M}x+\alpha)^2}{(\frac{M}{2}+x)^2},\\
		&g_y(x)=\frac{(1-\alpha)^2+(\frac{2\alpha}{M}x-1)^2}{(\frac{M}{2}-x)^2},
	\end{align}
	$\theta_0=\tan^{-1}(x_2)$ and $\theta_1=\tan^{-1}(\frac{M^2(3M^2-6M+2)x_1}{(M-2)(3M-2)(M^2-2M+2)})$ where $x_2$ and $x_1$ are the bigger and the smaller root of the following equation (solved for $x$), respectively
	\begin{align}
		&M^2(3M^2-6M+2)x^2-2M(3M^3-10M^2+12M-4)x\nonumber\\
		&+(M-2)(3M-2)(M^2-2M+2)=0. \label{equa}
	\end{align}
\end{theorem}
\begin{proof}
	Please see Appendix \ref{App2}.
\end{proof}
Using same procedure the following sets are the optimal start set for other ranges of $\theta$:

If $\frac{\pi}{2}<\theta\le\frac{3\pi}{4}-\theta_0$,
\begin{align}
	\mathcal{S}_3=\{(0,0),(\frac{M}{2}, \frac{M}{2}),(\frac{M}{2}, y_2)\},
\end{align} 
where
\begin{equation*}
	y_2 = \left\{
	\begin{array}{rl}
		\lfloor \frac{-M(2+\alpha)}{2\alpha}\rfloor & \emph{if } g_{2,y}(\lfloor \frac{-M(2+\alpha)}{2\alpha}\rfloor) \le g_{2,y}(\lceil \frac{-M(2+\alpha)}{2\alpha}\rceil)\\
		\lceil \frac{-M(2+\alpha)}{2\alpha}\rceil & \emph{if } g_{2,y}(\lfloor \frac{-M(2+\alpha)}{2\alpha}\rfloor) > g_{2,y}(\lceil \frac{-M(2+\alpha)}{2\alpha}\rceil),\\
	\end{array} \right.
\end{equation*}

If $\frac{3\pi}{4}-\theta_0<\theta<\frac{3\pi}{4}+\theta_1$,
\begin{align}
	\mathcal{S}_3=\{(0,0),(\frac{M}{2}, \frac{M}{2}-1),(\frac{M}{2}-1, \frac{M}{2})\},
\end{align} 

If $\frac{3\pi}{4}+\theta_1\le\theta<\pi$,
\begin{align}
	\mathcal{S}_3=\{(0,0),(\frac{M}{2}, \frac{M}{2}),(x_2, \frac{M}{2})\} \label{1opt},
\end{align} 
where
\begin{equation*}
	x_2\hspace{-0.5mm} = \hspace{-0.5mm}\left\{\hspace{-2mm}
	\begin{array}{rl}
	\lfloor \frac{-M(1+2\alpha)}{2}\rfloor &\hspace{-1.3mm}\emph{if } g_{2,x}(\lfloor \frac{-M(1+2\alpha)}{2}\rfloor) \le g_{2,x}(\lceil \frac{-M(1+2\alpha)}{2}\rceil)\vspace{1mm}\\
	\lceil \frac{-M(1+2\alpha)}{2}\rceil &\hspace{-1.3mm}\emph{if } g_{2,x}(\lfloor \frac{-M(1+2\alpha)}{2}\rfloor) > g_{2,x}(\lceil \frac{-M(1+2\alpha)}{2}\rceil),\\
	\end{array} \right.
\end{equation*}
\begin{align}
	&g_{2,x}(x)=\frac{(1+\alpha)^2+(\frac{2}{M}x+\alpha)^2}{(\frac{M}{2}-x)^2},\\
	&g_{2,y}(x)=\frac{(1+\alpha)^2+(\frac{2\alpha}{M}x+1)^2}{(\frac{M}{2}-x)^2}.
\end{align}

For $\pi<\theta\le\frac{5\pi}{4}-\theta_0$, $\frac{5\pi}{4}-\theta_0<\theta<\frac{5\pi}{4}+\theta_1$, $\frac{5\pi}{4}+\theta_1\le\theta<\frac{3\pi}{2}$, $\frac{3\pi}{2}<\theta\le\frac{7\pi}{4}-\theta_0$, $\frac{7\pi}{4}-\theta_0<\theta<\frac{7\pi}{4}+\theta_1$ and $\frac{7\pi}{4}+\theta_1\le\theta<2\pi$ the optimal choices are exactly same as the ones for $0<\theta\le\frac{\pi}{4}-\theta_0$, $\frac{\pi}{4}-\theta_0<\theta<\frac{\pi}{4}+\theta_1$, $\frac{\pi}{4}+\theta_1\le\theta<\frac{\pi}{2}$, $\frac{\pi}{2}<\theta\le\frac{3\pi}{4}-\theta_0$, $\frac{3\pi}{4}-\theta_0<\theta<\frac{3\pi}{4}+\theta_1$ and $\frac{3\pi}{4}+\theta_1\le\theta<\pi$, respectively. 

By virtue of Theorem~\ref{T2}, the choice of the first three antennas for any $M$ is reduced to only two sets. For example, suppose we have an array of $81$ antennas. In that case, the total number of possible sets for the start point is $\binom{80}{2}=3160$ cases. Such a search pool can prevent antenna selection in real-time, but this problem is solved with the help of Theorem~\ref{T2}. Now that we have an optimal starting point for the greedy algorithm, by using the characteristics of the $CRLB_\theta$ formulation, it is possible to reduce the search area of the greedy algorithm to half. 
\begin{corollary} \label{cor2}
	Every antenna selected in the Main Stage, either for the start set or by the greedy algorithm, has the same contribution in the $CRLB_\theta$ as its reciprocal counterpart w.r.t the reference point and can be replaced with it.
\end{corollary}
\begin{proof}
	Consider that in the $n$th step of the greedy algorithm if $(-x_1,-y_1)$ is added instead of $(x_1,y_1)$. By noting the formulation of $U(s)$, at the nominator all the elements of both $\tilde{\boldsymbol{\Sigma}}_1$ and $\tilde{\boldsymbol{\Sigma}}_2$ are squared before addition. So, all $x$ and $y$ elements are squared and changing the sign of those elements will not affect the nominator. In the denominator, again the $tr(\tilde{\boldsymbol{\Sigma}}_1^2)tr(\tilde{\boldsymbol{\Sigma}}_2^2)$ will not be affected due to the fact that their elements are squared. For $tr(\tilde{\boldsymbol{\Sigma}}_1\tilde{\boldsymbol{\Sigma}}_1)$, each $x$ is multiplied by its corresponding $y$, so if the sign of both of them are changed simultaneously, the final result will not change. 
\end{proof}

By virtue of the Corollary~\ref{cor2}, the search area of the greedy algorithm will be restricted only to half of the antenna array. This is because when the contribution of the antennas in the first half is calculated, the contribution of all of the antennas in the other half will be the same as the first one. This significantly reduces the computational complexity of the greedy algorithm, which is essential for real-time utilization. The complete procedure of the real-time antenna selection algorithm is presented in Table.~\ref{Table1}. 

Contrary to previous implementations of greedy algorithms for antenna selection, the real-time antenna selection algorithm presented in Table.1 uses the instantaneous $CRLB_\theta$ of a planar antenna array as its cost function. This feature is possible due to two main entries that are fed into the algorithm. First is the preliminary estimation ($\theta_p$) acquired in the algorithm's preliminary stage. Second, the optimal start point for every $\theta_p$ composed of three antennas. Calling the total number of antennas $N$, and noting that only the Main stage is going to be run in real-time, the algorithm at the $f$th stage has to search an antenna pool composed of $N-(F-3)$ antennas. So, the total number of iterations to select $F$ antennas is

\begin{equation}
	\sum_{f=1}^{F-3}(N-f)=(F-3)N-\frac{(F-3)(F-2)}{2}=(F-3)(N-\frac{F}{2}+1). \label{CCo}
\end{equation}

The highest order of complexity in \eqref{CCo} is $\mathcal{O}(NF)$. The presented algorithms in \cite{wang2014adaptive} and \cite{aboutanios2021online} have complexity order of $\mathcal{O}(N^3)$ and $\mathcal{O}(N^6)$, respectively. By noting $F<N$, we have $\mathcal{O}(NF)<\mathcal{O}(N^3)<\mathcal{O}(N^6)$, meaning that our algorithm has much lower computational complexity compared to these works. Also, methods in \cite{wang2014adaptive,aboutanios2021online} have an optimization inside their algorithm, where we only have a simple calculation of the $CRLB$ function that further reduces the computational complexity.

In certain applications where the AoA of a UT is being tracked like \cite{li2019massive,zhu2015tracking}, the preliminary stage of the antenna selection can be ignored. Suppose the frequency of tracking estimation is fast enough that the AoA of the UT is not hugely changed. In that case, based on the previous estimation of the UT's AoA, the system has a fair idea of the current AoA. So, the preliminary estimation stage can be removed, and $\theta_p$ can be replaced with the estimation of the previous AoA. This will help such applications implement real-time antenna selection with even lower computational complexity. 

\begin{remark}
	From \eqref{CRLB}, \eqref{ECRLB}, \eqref{p1} and the presented results in \cite{arash2021analysis}, it is evident that changing the channel model to the DMC model does not affect the antenna selection procedure. The real-time antenna selection can also be used in the DMC channels. Also, in the presence of inter-symbol interference (ISI) with independent channel coefficients and AoAs will not change the antenna selection part in \eqref{CRLB}, \eqref{ECRLB} and \eqref{p1}. So, the presented antenna selection methods can be implemented for any distribution of channel coefficients and/or in the presence of independent ISI.
	%From \eqref{CRLB}, \eqref{ECRLB} and \eqref{p1} it is evident that changing the distribution of channel coefficients to consider path-loss signals does not affect the configuration of selected antennas. Indeed, these changes will only change the $|h_d|^2$ coefficient in the aforementioned equations \cite{arash2021analysis}.
\end{remark}
\begin{table}
	\centering
	\begin{tabular}{|l|c|}
		\hline
		\textbf{get} $F_p$,\ $F$&\\
		\hline
		Select $F_p$ antennas based on Theorem\ref{T1}.& Preliminary\Tstrut\\
		Estimate $\theta_p$. & Stage\\
		\hline
		Using $\theta_p$, select $\mathcal{S}_3$ based on Theorem\ref{T2}.&\Tstrut\\
		$\mathcal{S}:=\mathcal{S}_3$&\\
		\textbf{For}  $f:=4 \hspace{1mm} \textbf{to} \hspace{1mm} F$\hspace{1mm}\textbf{do}&\\
		\hspace{6mm}\textbf{For}  $x:=0 \hspace{1mm} \textbf{to} \hspace{1mm} M/2$\hspace{1mm}\textbf{do}&\\
		\hspace{12mm}\textbf{For}  $y:=0 \hspace{1mm} \textbf{to} \hspace{1mm} M/2$\hspace{1mm}\textbf{do}&\\
		\hspace{18mm}\textbf{If}  $(x,y)\notin \mathcal{S} $\hspace{1mm} \textbf{do}&\\
		\hspace{24mm}$\mathcal{S}(f):=(x,y)$&Main\\
		\hspace{24mm}$\boldsymbol{V}(x,y):=CRLB_\theta(\mathcal{S})$&Stage \\
		\hspace{18mm}\textbf{EndIf}&\\
		\hspace{12mm}\textbf{EndFor}&\\
		\hspace{6mm}\textbf{EndFor}&\\
		\hspace{6mm}$\mathcal{S}(f):=\arg\displaystyle\min_{(x,y)}\boldsymbol{V}$&\\
		\textbf{EndFor}&\\
		\textbf{return} $\mathcal{S}$&\\
		\hline
	\end{tabular}
	\caption{Real-time antenna selection algorithm.}
	\label{Table1}
\end{table}

\section{Numerical Results} \label{NumericalR}
In this section, we test the presented theorems by Monte-Carlo simulations. As a benchmark, the performance of real-time antenna selection is compared with the expected antenna selection method. In the expected method, only the preliminary stage (presented in section\ref{PS1}), of the proposed antenna selection is used (also related to our work previous work \cite{arash2021analysis}). Also, we compare both selection methods with a strategy in which the preliminary stage is done, but the start point of the greedy algorithm is randomly selected, and $\theta_p$ is only used in the greedy algorithm's cost function (i.e. Theorem\ref{T2} is not used)\footnote{Unlike \cite{he2021reconfigurable,bhattacharya2021selection,zhang2020antenna}, we do not compare our results with total random selection method (in which all of the antennas are selected randomly) as it performs so poorly that it distorts the scale of plots. One can better appreciate small differences in different selection methods in the current view.}. Unless otherwise stated, in the following figures, $\frac{d}{\lambda}=0.5$, $|h_d|^2=1$, $\rho=0dB$ and $\varphi=\frac{\pi}{3}$.

\begin{figure}[t]
	\centering
	\includegraphics[width=0.4\textwidth]{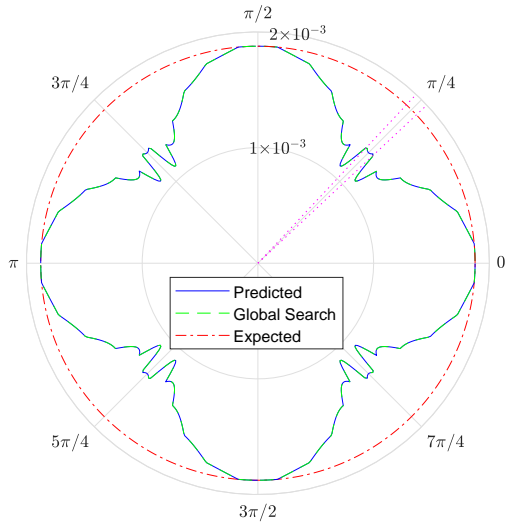}
	\caption{$CRLB_\theta$ versus $\theta$ for $F=3$ and different selection methods.}
	\label{ThreeAntenna}
\end{figure}

Fig.~\ref{ThreeAntenna} shows the $CRLB_\theta$ for all possible AoAs when only three antennas are selected, using different methods for $M=6$. The red curve is $CRLB_\theta$ when three antennas are selected to minimize $\mathbb{E}_\theta\{CRLB_\theta\}$. The blue and green curves result from selecting antennas to minimize instantaneous $CRLB_\theta$, using Theorem \ref{T2} and global search, respectively. Moreover, the dotted lines show $\theta_0=43.1352^\circ$ and $\theta_1=46.8648^\circ$, where the optimal choices change according to \eqref{pi/4} and \eqref{equa}. It is seen that Theorem~\ref{T2} is perfectly capable of predicting the optimal set for the first three antennas.

\begin{figure*}[t!]
	\centering
	\begin{subfigure}[t]{0.4\textwidth}
		\includegraphics[width=\textwidth]{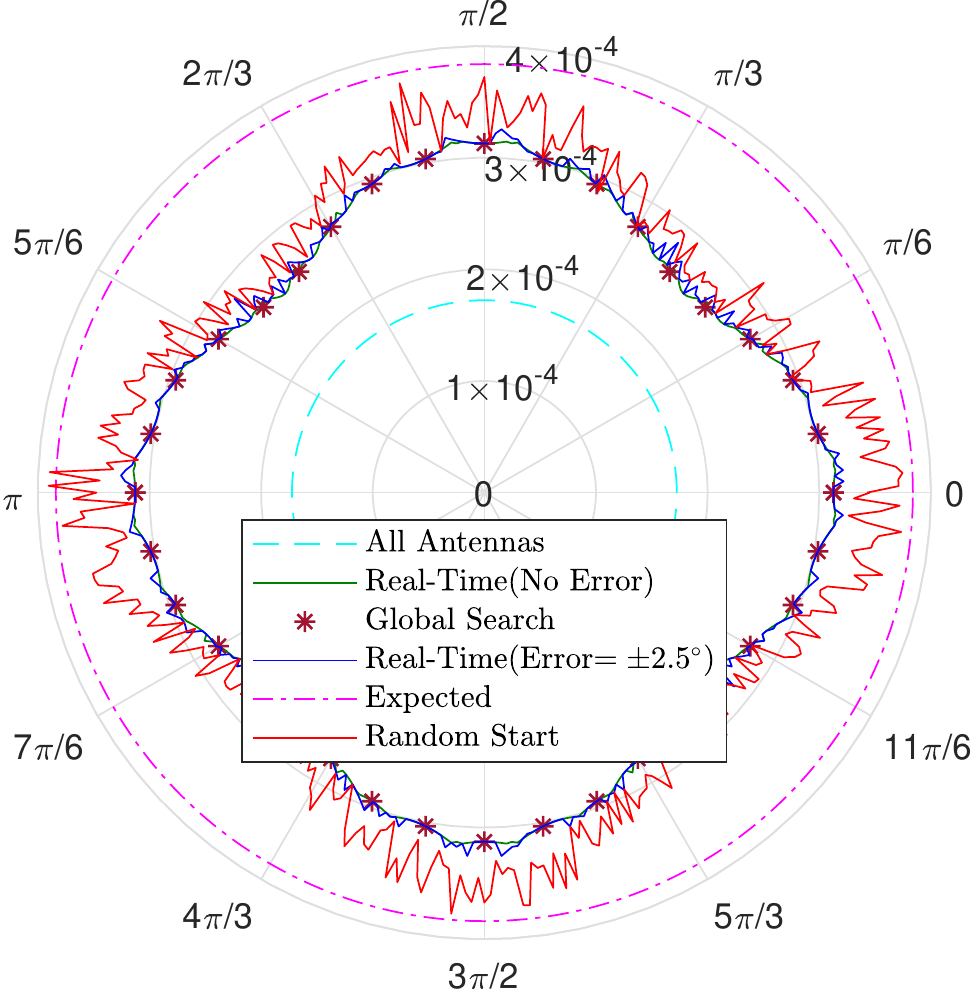}
		\caption{$F=13$ and $M=6$.} \label{TotalCRLB}
	\end{subfigure}
	~
	\begin{subfigure}[t]{0.4\textwidth}
		\includegraphics[width=\textwidth]{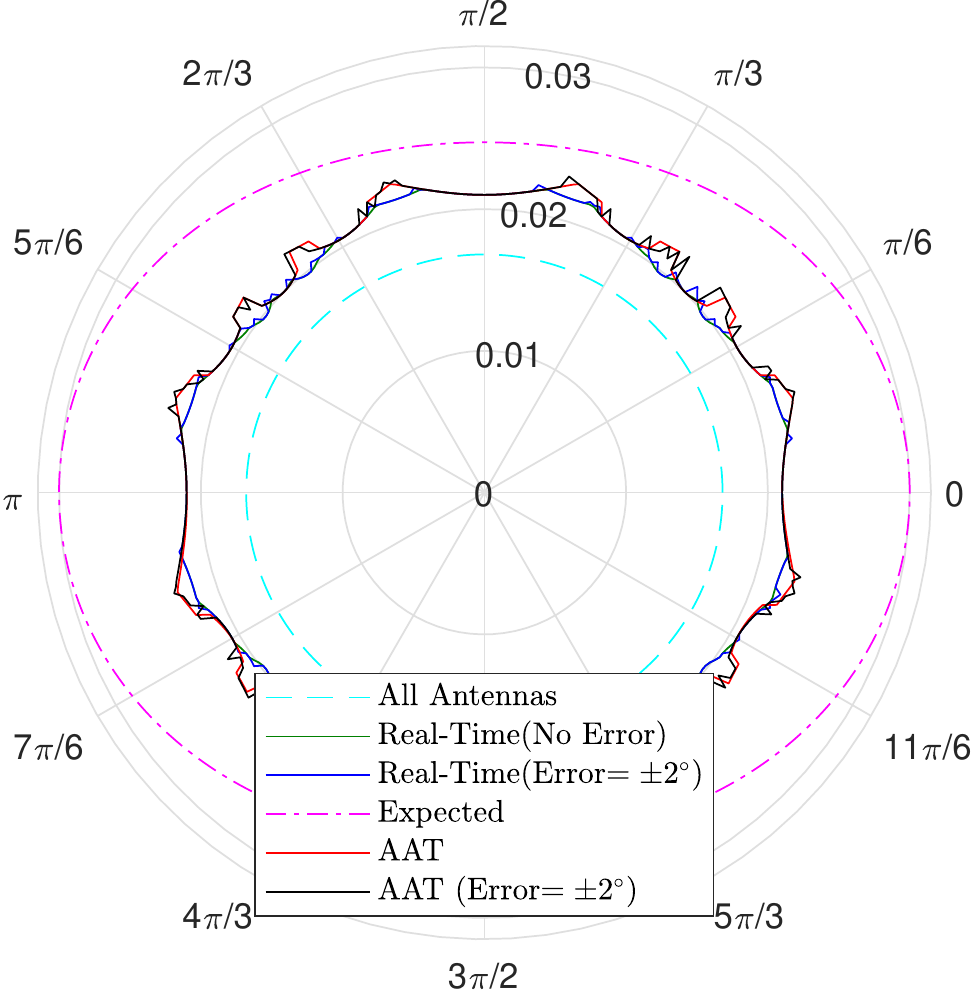}
		\caption{$F=11$ and $M=4$.} \label{comp}
	\end{subfigure}
	\caption{$CRLB_\theta$ versus $\theta$ for different selection strategies and scenarios.}
\end{figure*}\label{Compare}

Fig.~\ref{TotalCRLB} indicates the resulted $CRLB_\theta$ versus $\theta$ for different antenna selection strategies and scenarios. In this figure, $M=6$ and $F=13$. The magenta curve results when antennas are selected to minimize $\mathbb{E}_\theta\{CRLB_\theta\}$. The cyan curve is for the case when all of the available antennas are used ($F=(M+1)^2$). The green curve results from the real-time greedy method, when AoA is perfectly known, i.e., $\theta_p=$AoA. It is seen that the real-time antenna selection method secures between $52\%$ and $68\%$ of total performance (depending on the value of $\theta$), only by using a quarter of the total available antennas. Moreover, the expected antenna selection method achieves $\%45$ of total performance. The star markers ($*$) show the $CRLB_\theta$ that is obtained through global search, i.e., the global minimum for $F=13$. It is seen that the greedy algorithm obtains a very close performance to global search with much less computational complexity, indicating that the greedy algorithm with an optimal start point is a reasonable choice for antenna selection. In the blue curve, constant noise is added to the $\theta_p$ to show how the selection strategy performs when $\theta_p\ne$ AoA. The noise value is $2.5^\circ$ with a random positive or negative sign, i.e., $\theta_p=$AoA$\pm2.5^\circ$. It should be noted that the $CRLB_\theta$ is in the order of $3\times10^{-4}$ ($\approx0.01^\circ$), and the added error is more than two orders of magnitude higher than $CRLB_\theta$. It is seen that even with this amount of error, the selection process is performing so close to the optimal case. Also, this curve shows that different AoAs have different tolerances against error in $\theta_p$. It should be noted that an error in $\theta_p$ affects both parts of the real-time selection algorithm, the start set and the greedy part. This amount of error tolerance can motivate one to create a look-up table for the real-time antenna selection by selecting a set for every $2.5^\circ$ (or any other value) and using a proper set according to the value of $\theta_p$ to extremely simplify the antenna selection procedure. The red curve with a random start set is plotted to see the effects of the optimal start set. In this curve, the optimal start set proposed by Theorem\ref{T2} is not used. Instead, two random antennas are selected in addition to the reference point. The greedy part is the same as the blue curve, and $\theta_p$ is used in the greedy part of the red curve. It is seen that, in general, $CRLB_\theta$ deteriorates when the optimal start point is not used, sometimes even worse than the expected method. This highlights the importance of the optimal start point, presented in Theorem\ref{T2}.

In Fig.\ref{comp}, our proposed method is compared with another antenna selection method proposed in \cite{wang2014adaptive}. In \cite{wang2014adaptive}, a Dinklebach type algorithm is used in CVX to minimize the $CRLB$. This selection method is named \emph{AAT}. AAT is compared in two scenarios in Fig.~\ref{comp}, first actual $\theta$ is fed into the algorithm, and second, $2^\circ$ of error is added when feeding the algorithm. Our proposed real-time method is also simulated in the same scenarios. In all the cases, $F=11$. It is seen that with the same inputs, when actual $\theta$ is used, the real-time algorithm is always better. Also, when there is an error, the real-time algorithm performs better in most points. Our algorithm obtains this performance with much less computational complexity (resulting in lower run time) and the possibility of being implemented in real-world scenarios.

\begin{figure}[t]
	\centering
	\includegraphics[width=0.45\textwidth]{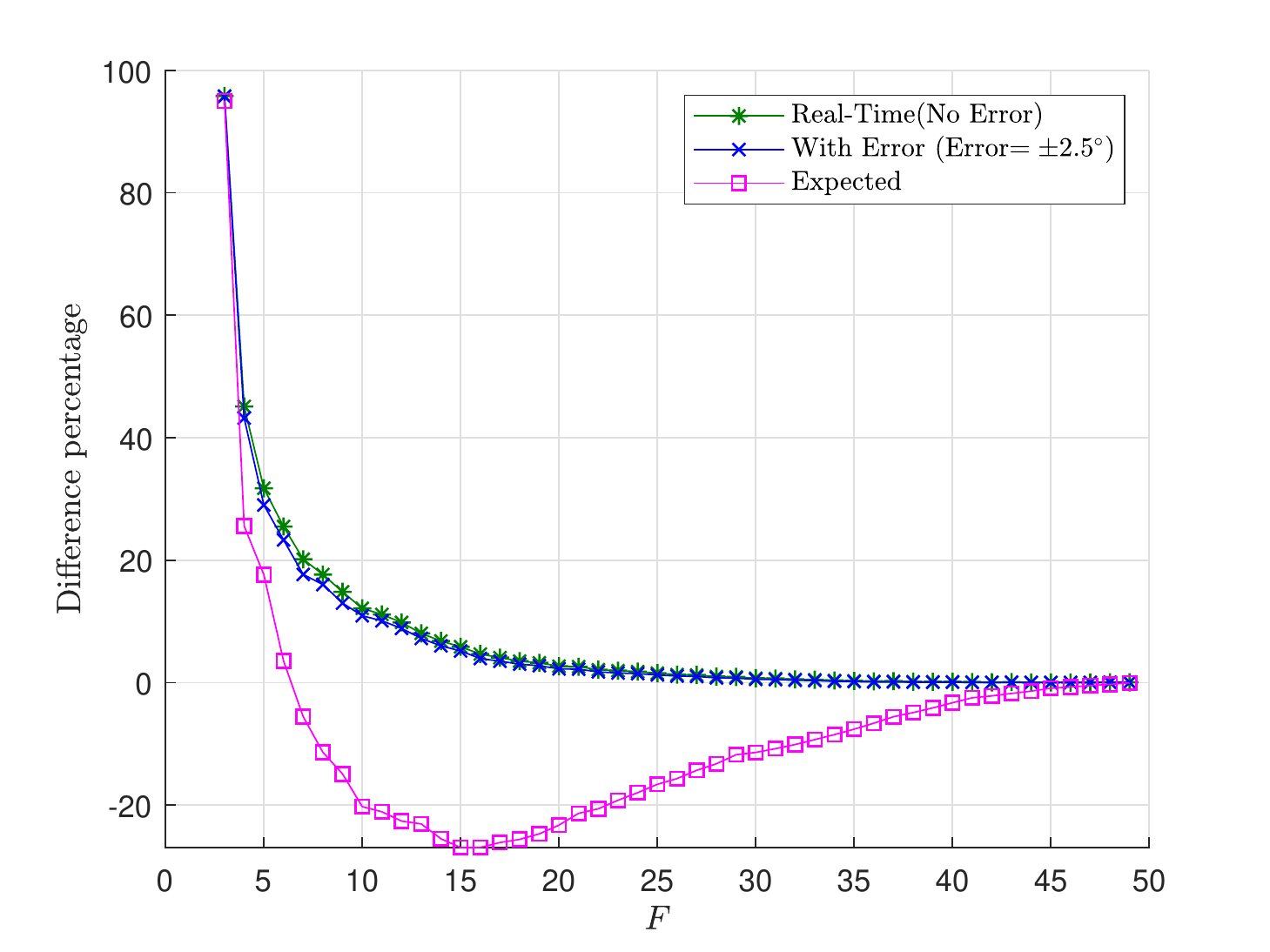}
	\caption{Performance improvement percentage of different methods relative to real-time method with random start point.}
	\label{Percent}
\end{figure}

Fig.~\ref{Percent} shows the relative advantage of real-time and expected antenna selection methods to the selection method in which only the start point is selected randomly, i.e., $\theta_p$ is only used in the greedy part. For all curves, each point is calculated as $\frac{\sum_{\theta=0}^{2\pi}([CRLB_\theta]_R-[CRLB_\theta]_A)}{N\sum_{\theta=0}^{2\pi}[CRLB_\theta]_R}$, where $[CRLB_\theta]_A$ means the $CRLB_\theta$ resulted by using method $A$, and $A$ can be real-time (either with or without error) or expected method, $[CRLB_\theta]_R$ means the $CRLB_\theta$ resulted by using real-time method with random start point and $N$ is total number of angles that have been summed. This figure aims to show that for every $F$, how using proposed methods, namely real-time with optimal start point, increases the performance w.r.t the case in which the real-time method is utilized but with the random start point. It is seen that when $F$ is relatively small, by using the presented optimal start set, the performance increases significantly. As $F$ grows, the difference in performance decreases. This is because as more and more antennas are selected, even if the start set is selected randomly, other antennas that are selected in the greedy section by the random algorithm will make up a considerable portion of the total selected antennas. So the contribution of two optimally selected antennas for the start set will decrease. Therefore, as more antennas are selected, the importance of the optimal start set reduces because antennas with high contributions will be selected one way or another. Another point of Fig.~\ref{Percent} is the behavior of the curve for the expected antenna selection method. First, it has a positive relative performance. Then it becomes negative, meaning that for certain $F$ expected method is better than the real-time with a random start point. As $F$ increases, the real-time approach, even with a random start point, is better than the expected method.
\begin{figure*}[t]
	\centering
	\begin{subfigure}[t]{0.45\textwidth}
		\centering
		\includegraphics[width=\textwidth]{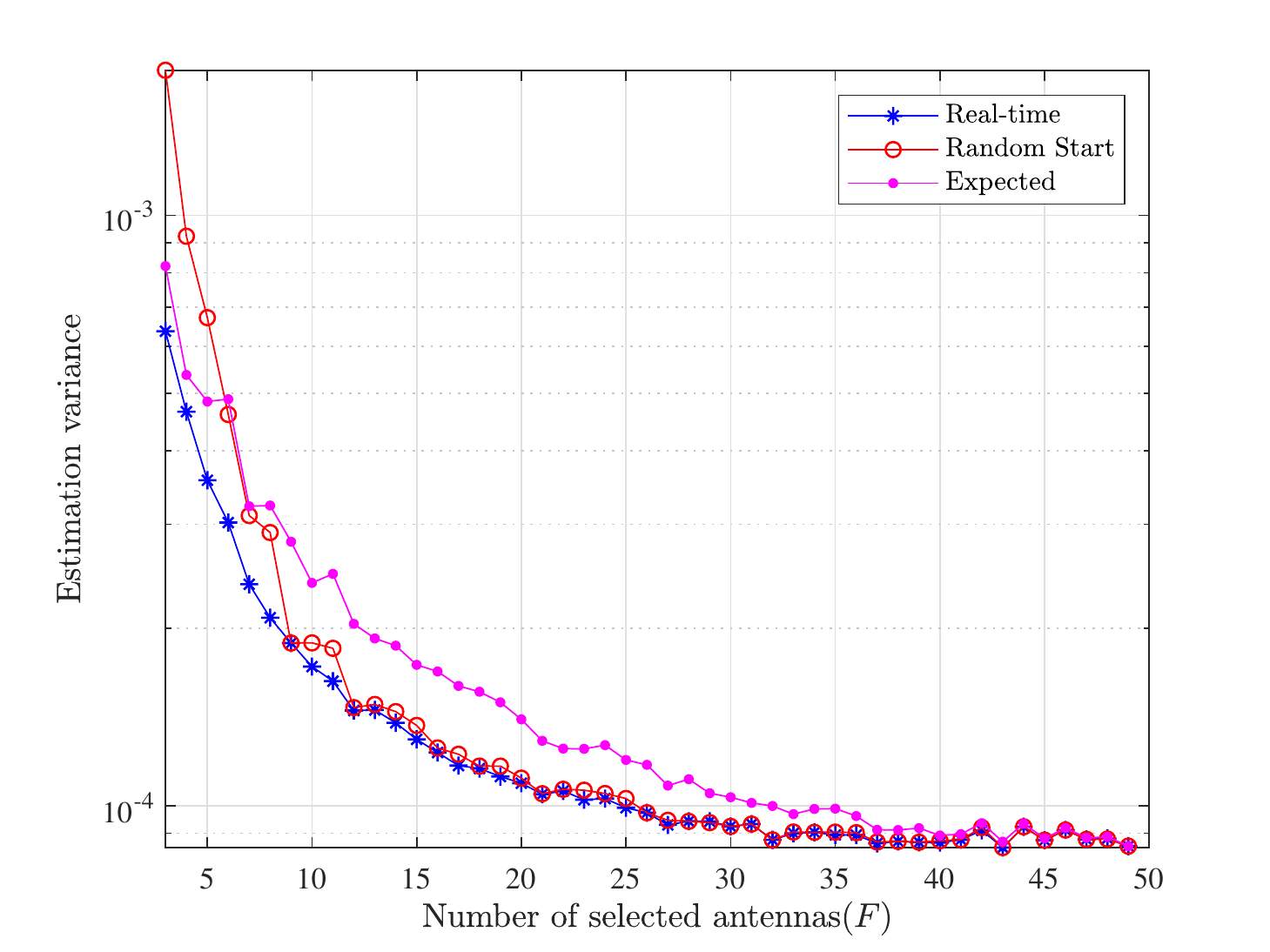}
		\caption{Variance of ML estimator for different number of antennas.} \label{ML}
	\end{subfigure}
	\hfill
	\centering
	\begin{subfigure}[t]{0.45\textwidth}
		\centering
		\includegraphics[width=\textwidth]{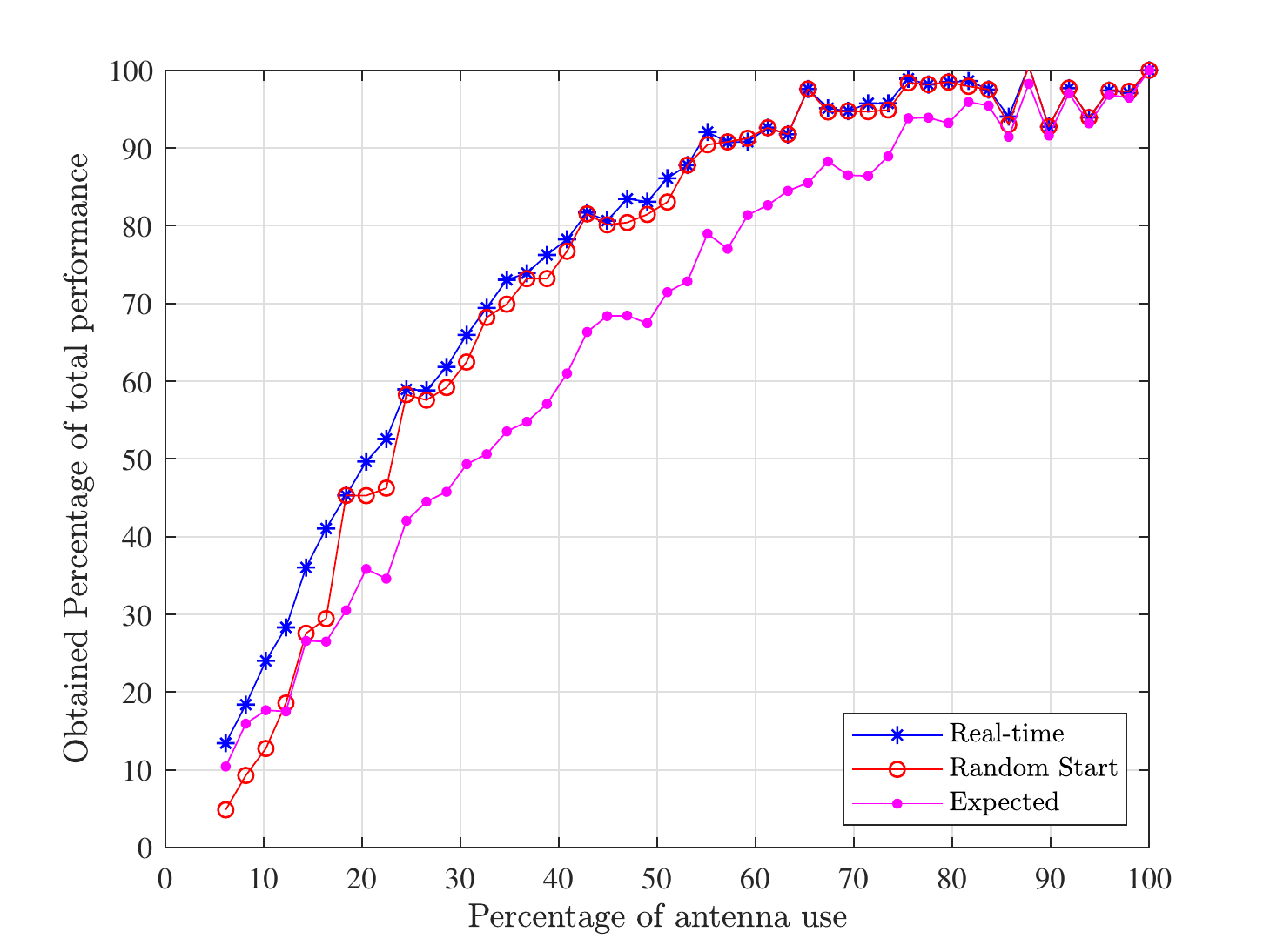}
		\caption{Percentage of obtained performance versus percentage of used antennas.} \label{eff}
	\end{subfigure}
	\caption{Performance analysis of ML estimator with different antenna selection methods.}\label{estimate}
\end{figure*}

In Fig.~\ref{estimate}, antenna selection is applied to the Maximum-likelihood estimator. In this figure, $\theta=\pi/6$, $\varphi=\pi/3$, $F_p=4$ and, $M=6$. Fig.~\ref{ML} indicates the estimation variance for different numbers of selected antennas. As predicted by Fig.~\ref{Percent}, soon after small values of $F$, real-time antenna selection performs better than the expected method. Also, as $F$ increases, real-time selection with optimal and random start becomes closer. In Fig.~\ref{eff}, the percentage of obtained performance versus the percentage of utilized antennas are plotted to show the effectiveness of each antenna selection method. The real-time antenna selection obtains more than $80\%$ of total performance only by using half of the available antennas, showing the effectiveness of antenna selection in improving both hardware and energy efficiency.

\begin{figure}[t]
	\hspace*{+3.52cm}   
	\includegraphics[width=0.5\textwidth]{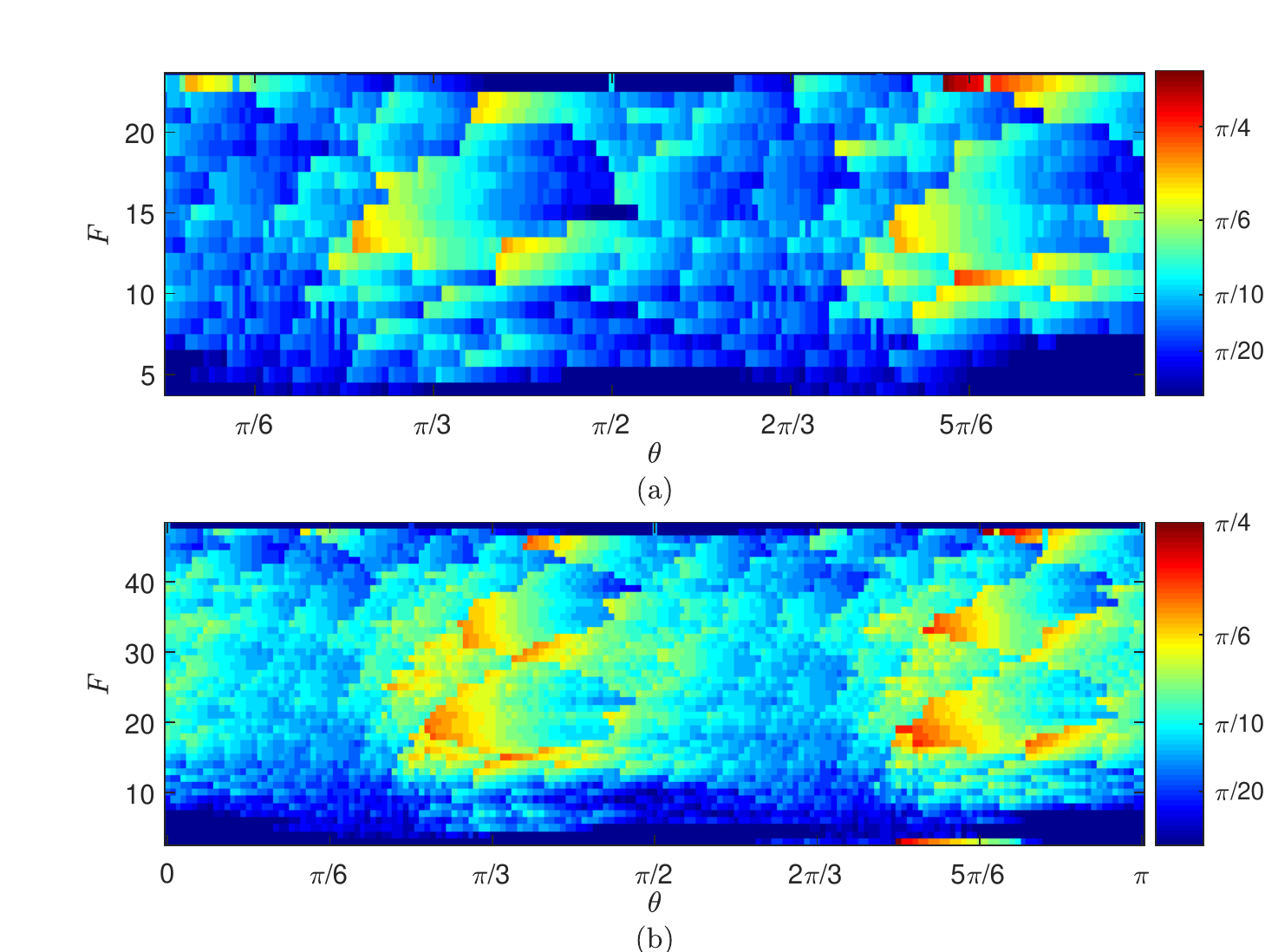}
	\caption{Error tolerance of real-time antenna selection strategy for: (a) $M=4$ and (b) $M=6$.}\label{ERT}
\end{figure}

Fig.~\ref{ERT} is plotted to further examine how the main real-time antenna selection method (the one that takes advantage of the optimal start set) performs compared to the expected method. This figure represents how error tolerant the real-time antenna selection is, compared with the expected antenna selection method. In this figure, the x-axis is the AoA, and the y-axis is the number of selected antennas. For each point in this 2D plane, the color indicates how much error the real-time algorithm can bear in the preliminary stage and still have lower $CRLB_\theta$ than the antenna selection method that minimizes expected $CRLB_\theta$. A deep blue color means lower error tolerance, and the more the color of a point becomes reddish, the higher the error can be tolerated by the real-time algorithm. It is seen that starting after $\frac{\pi}{3}$ up to around $\frac{\pi}{2}$, between $5\le F\le20$ for $M=4$ and between $10\le F\le40$ for $M=6$, the real-time algorithm shows its highest resistance. Moreover, increasing the number of antennas makes the real-time algorithm more error tolerant. This means that the real-time selection method performs very well when $F$ is around half of the $M$, i.e., for relatively moderate values of $F$. Also, it should be noted that the reason for the low error tolerance in higher values of $F$ is that as most of the antennas are already selected by both selection strategies, the $CRLB_\theta$ is reaching its final value (that is using $(M+1)^2$ antennas) and then even selecting one antenna with lower priority results in a very small but higher value for $CRLB_\theta$, so error tolerance is low. However, despite lower tolerance, the value of resulted $CRLB_\theta$ is very close for both strategies. 
\begin{figure}[t]
	\centering
	\includegraphics[width=0.45\textwidth]{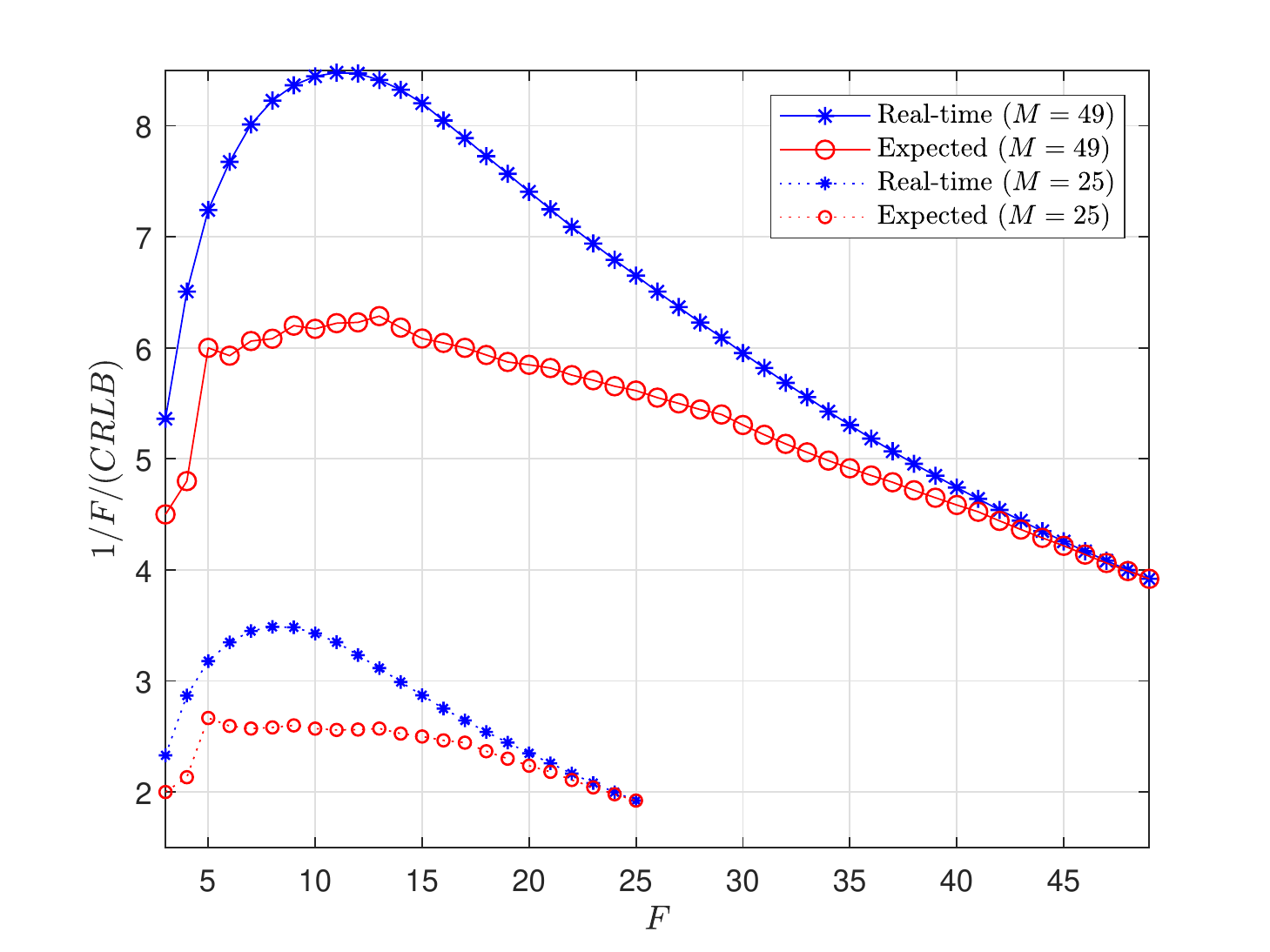}
	\caption{The inverse of the product of $CRLB$ and $F$, versus $F$.}
	\label{Eff}
\end{figure}

In Fig.~\ref{Eff}, the inverse product of $F$ and $CRLB_\theta$ is plotted for $M=6$ and $M=4$ to show how efficiently each selection strategy is using its selected antennas. This can be seen as a basic study and can be modified for any efficiency study, like the energy efficiency of AoA estimation introduced in our previous work \cite{Arash2020LE}. In this figure, in both real-time and expected selection methods for each $F$, the average value for all possible $\theta$s are plotted, i.e. $\frac{1}{N}\sum_{\theta=0}^{2\pi}CRLB_\theta$ where $N$ is total number of $\theta$s that are summed. It is seen that the real-time selection method always obtains higher efficiency in terms of antenna usage. As expected from Fig.~\ref{ERT}, the region where real-time selection has the highest efficiency difference w.r.t the expected antenna selection corresponds to the same region where real-time selection also has higher error tolerance. Also, noting that the endpoints of the curves indicate the efficiency of antenna usage without antenna selection, significantly higher efficiencies can be obtained by making use of antenna selection.

\begin{figure}[t]
	\centering
	\includegraphics[width=0.45\textwidth]{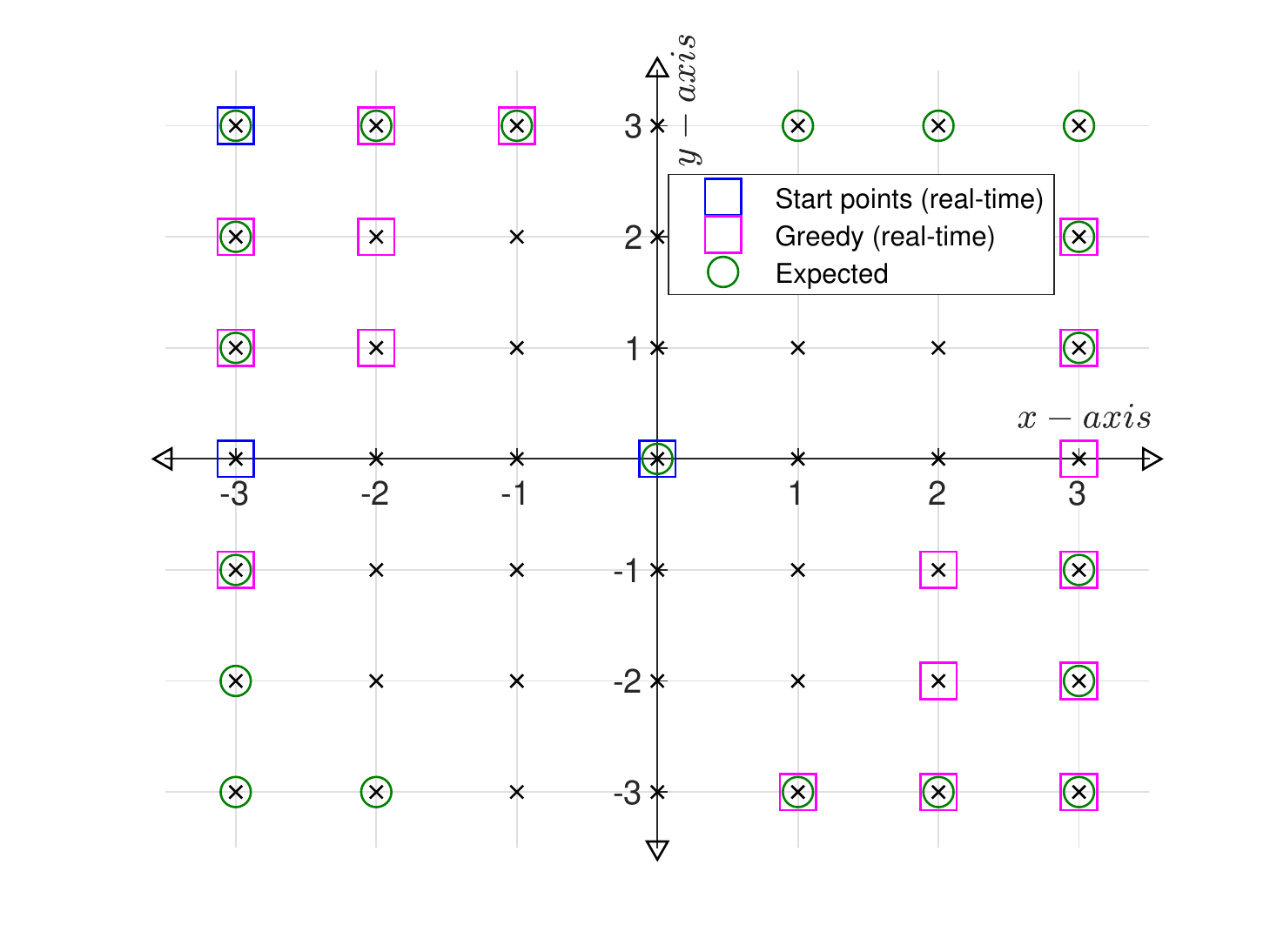}
	\caption{Selected antennas by both real-time and expected algorithms for $F=20$ and $\theta=\frac{\pi}{3}$.}
	\label{SA}
\end{figure}

Finally, in Fig.~\ref{SA} an example of the selected antennas by the presented algorithms for $M=6$, $F=20$ and $\theta=\frac{\pi}{3}$ is plotted. In this figure, square markers show the selected antennas by the real-time antenna selection (without error). The blue squares indicate start points for the greedy algorithm, and the greedy algorithm selected magenta ones. Also, green circles demonstrate the selected antennas by the expected algorithm. We observe that the greedy algorithm creates two separate subarrays to estimate AoA. This result that several smaller subarrays perform better than one big array is also reported for large intelligent surfaces in \cite{hu2018beyond} and Massive MIMO systems in \cite{arash2021analysis}. Our results for both expected and real-time antenna selection methods are in accordance with these results. 
\section{Conclusion}
This paper studied antenna selection for AoA estimation for a square planar antenna array. We proposed two CRLB based antenna selection methods for AoA estimation. In the first method, motivated by our previous studies, the goal is to minimize the expected value of the $CRLB_\theta$. The optimal strategy for antenna selection in this scenario is presented.

Next, a two-step real-time antenna selection method is developed using the first introduced antenna selection method. This method selects an initial set of antennas to minimize the expected $CRLB_\theta$. Using this set, a fast and crude estimation of AoA is done. Then the estimated AoA is used in a greedy algorithm whose objective function is to minimize instantaneous $CRLB_\theta$. The optimal start point of the algorithm is presented to make the algorithm fast enough (so it can be run in real-time). Moreover, a corollary that reduces the search space to half the available antennas is presented to reduce the algorithm's computational complexity. 

Numerical results show a considerable gain in performance when either one of the methods is employed. Also, it is shown that a greedy algorithm with an optimal starting point obtains quite close performance to global search. Furthermore, both presented algorithms' error tolerance and efficiency of antenna usage are compared with one another and with the real-time method, whose start point is random. It is seen that for a moderate number of active antennas, the real-time antenna selection (with optimal start point) has its highest error tolerance and efficiency w.r.t the antenna selection method that aims to minimize expected $CRLB_\theta$. The generalization of the real-time algorithm to the multi-user scenario is a direction for future works.
\section{acknowledgment}
This work is supported by F.R.S.-FNRS under the EOS research project MUSEWINET (EOS project 30452698).

\appendices
\section{Proof of Theorem \ref{T1}} \label{App1}
The process of proof is to show that for every step, the presented set in Theorem\ref{T1} obtains the lowest possible value for $U(\mathcal{S})$. We show this for the first two steps, i.e. for $F_p\le5$ and $5<F_p\le13$. The rest of the proof is exactly the same as these steps and can be simply repeated until any arbitrary $F_p$. 

For $F_p=5$, selecting $S_{p_1}$ will result in $U(\mathcal{S}_{p_1})=\frac{2}{M^2}$. Now we show that any other set than $S_{p_1}$, results in higher value for $U(\mathcal{S})$. Assume any $\bar{S}_{p_1}=\{(0,0),(x_1,y_1),(x_2,y_2),(x_3,y_3),(x_4,y_4)\}$ to be a set of antennas which at least has one different antenna from $S_{p_1}$, we have
\begin{align}
	U(\mathcal{\bar{S}}_{p_1})=\frac{\sum_{i=1}^{5}(x_i^2+y_i^2)}{(\sum_{i=1}^{5}x_i^2)(\sum_{i=1}^{5}y_i^2)-(\sum_{i=1}^{5}x_iy_i)^2}. \label{uspb}
\end{align}	
Because at least one antenna in $\mathcal{\bar{S}}_{p_1}$ has at least one different coordinate from the antennas in the $\mathcal{S}_{p_1}$, so at least either $\sum_{i=1}^{5}x_i^2<M^2$ or $\sum_{i=1}^{5}y_i^2<M^2$, so
\begin{align}
	(M^2-\sum_{i=1}^{5}y_i^2)\sum_{i=1}^{5}x_i^2+(M^2-\sum_{i=1}^{5}x_i^2)\sum_{i=1}^{5}y_i^2&=M^2\sum_{i=1}^{5}(x_i^2+y_i^2)-2(\sum_{i=1}^{5}x_i^2)(\sum_{i=1}^{5}y_i^2)>0\nonumber\\&\ge-2(\sum_{i=1}^{5}x_iy_i)^2. \label{ineq1}
\end{align}	
Therefore
\begin{align}
	M^2\sum_{i=1}^{5}x_i^2+y_i^2>2((\sum_{i=1}^{5}x_i^2)(\sum_{i=1}^{5}y_i^2)-(\sum_{i=1}^{5}x_iy_i)^2).
\end{align}	
By noting that $(\sum_{i=1}^{5}x_i^2)(\sum_{i=1}^{5}y_i^2)-(\sum_{i=1}^{5}x_iy_i)^2>0$ due to Cauchy-Schwarz inequality, we obtain
\begin{align}
	\frac{\sum_{i=1}^{5}(x_i^2+y_i^2)}{(\sum_{i=1}^{5}x_i^2)(\sum_{i=1}^{5}y_i^2)-(\sum_{i=1}^{5}x_iy_i)^2}>\frac{2}{M^2}. \label{ineq31}
\end{align}	
When, $3<F_p<5$, all of greater signs ($>$) in \eqref{ineq1}-\eqref{ineq31} will be replaced with the greater or equal sign ($\ge$), with equality happening for any two subsets of $\mathcal{S}_{p_1}$. 

For $F_p=13$, the antennas set of $\mathcal{S}_{p_2}$ will result in $U(\mathcal{S}_{p_1}\bigcup\mathcal{S}_{p_2})=\frac{2}{2M^2+4(\frac{M}{2}-1)^2}$. For this set, $\sum_{i=1}^{5}x_i^2=\sum_{i=1}^{5}y_i^2=2M^2+4(\frac{M}{2}-1)^2$. We call $\mathcal{\bar{S}}_{p_2}$ a set in which at least one of the coordinates of one antenna is different from the $\mathcal{S}_{p_1}\bigcup\mathcal{S}_{p_2}$. In this case, depending on the fact that which coordinate is different, three scenarios will happen. First scenario is that either $\sum_{i=1}^{5}x_i^2$ or $\sum_{i=1}^{5}y_i^2$ is reduced and the other one remains constant. Second scenario is when both of them are reduced and the third scenario is when one is increased and the other one is reduced. It should be noted that no other choice for first $13$ antenna other than antennas in $\mathcal{S}_{p_2}$ can simultaneously increase both $\sum_{i=1}^{5}x_i^2$ and $\sum_{i=1}^{5}y_i^2$. This is because expect the reference point, all of antennas in $\mathcal{S}_{p_2}$ are the ones whose $x_i$s and $y_i$s are simultaneously maximized. 

In the aforementioned scenarios, the values of $\sum_{i=1}^{5}x_i^2$ and $\sum_{i=1}^{5}y_i^2$ for $\mathcal{S}_{p_2}$ will change to 
\begin{align}
	&\sum_{i\in \mathcal{\bar{S}}_{p_2}}x_i^2=2M^2+4(\frac{M}{2}-1)^2-a,\nonumber\\
	&\sum_{i \in \mathcal{\bar{S}}_{p_2}}y_i^2=2M^2+4(\frac{M}{2}-1)^2-b,
\end{align}
We have
\begin{align}
	&(2M^2+4(\frac{M}{2}-1)^2)(\sum_{i\in \mathcal{\bar{S}}_{p_2}}x_i^2+y_i^2)-2(\sum_{i\in \mathcal{\bar{S}}_{p_2}}x_i^2)(\sum_{i\in \mathcal{\bar{S}}_{p_2}}y_i^2)\nonumber\\&=
	(2M^2+4(\frac{M}{2}-1)^2)(\sum_{i\in \mathcal{{S}}_{p_2}}x_i^2+y_i^2-a-b)-2(\sum_{i\in \mathcal{{S}}_{p_2}}x_i^2-a)(\sum_{i\in \mathcal{{S}}_{p_2}}y_i^2-b)\nonumber\\&=(2M^2+4(\frac{M}{2}-1)^2)(\sum_{i\in \mathcal{{S}}_{p_2}}x_i^2+y_i^2)-2(\sum_{i\in \mathcal{{S}}_{p_2}}x_i^2)(\sum_{i\in \mathcal{{S}}_{p_2}}y_i^2)+(2M^2\nonumber\\&+4(\frac{M}{2}-1)^2)(a+b)-2ab\nonumber\\&=\underbrace{\left(2M^2+4(\frac{M}{2}-1)^2-\sum_{i\in \mathcal{{S}}_{p_2}}y_i^2\right)}_{=0}\sum_{i\in \mathcal{{S}}_{p_2}}x_i^2+\underbrace{\left(2M^2+4(\frac{M}{2}-1)^2-\sum_{i\in \mathcal{{S}}_{p_2}}x_i^2\right)}_{=0}\sum_{i\in \mathcal{{S}}_{p_2}}y_i^2\nonumber\\&+(2M^2+4(\frac{M}{2}-1)^2)(a+b)-2ab. \label{ab}%\nonumber\\&=(2M^2+4(\frac{M}{2}-1)^2-b)a+(2M^2+4(\frac{M}{2}-1)^2-a)b, 
	%(2M^2+4(\frac{M}{2}-1)-\sum_{i=1}^{13}y_i^2)^2\sum_{i=1}^{5}x_i^2\nonumber\\&+(2M^2+4(\frac{M}{2}-1)-\sum_{i=1}^{13}x_i^2)^2\sum_{i=1}^{5}y_i^2>0>	-(\sum_{i=1}^{5}x_iy_i)^2
\end{align}
Now, \eqref{ab} is a linear equation w.r.t $a$ and $b$. Its derivative w.r.t $a$ is
\begin{align}
	2M^2+4(\frac{M}{2}-1)^2-2b, \label{dera}
\end{align}
and \eqref{dera} is always positive because the maximum value of $b$ is $4(\frac{M^2}{4})$. So, the slope of \eqref{ab} is always positive. The most negative value of $a$ happens when all of four antennas with $x=\frac{M}{2}-1$ in $\mathcal{S}_{p_1}$ are replaced with $\frac{M}{2}$ in $\bar{\mathcal{S}}_{p_1}$. This replacement will result in $a=4((\frac{M}{2}-1)^2-(\frac{M}{2})^2)$ and $b=4((\frac{M}{2})^2-(\frac{M}{2}-2)^2)$. Therefore, $a+b>0$ in the most negative value of $a$. So, $a$ starts from a positive value with a positive slope, meaning that \eqref{ab} is positive for every possible value of $a$. Same thing can be shown w.r.t $b$. Therefore, 
\begin{align}
	&(2M^2+4(\frac{M}{2}-1)^2)(\sum_{i\in \mathcal{\bar{S}}_{p_2}}x_i^2+y_i^2)-2(\sum_{i\in \mathcal{\bar{S}}_{p_2}}x_i^2)(\sum_{i\in \mathcal{\bar{S}}_{p_2}}y_i^2)\nonumber\\&=(2M^2+4(\frac{M}{2}-1)^2-b)a+(2M^2+4(\frac{M}{2}-1)^2-a)b>0\ge-2\left(\sum_{i\in \mathcal{\bar{S}}_{p_2}}x_iy_i\right)^2. \label{29}
\end{align}
So, for any $\mathcal{\bar{S}}_{p_2}$ we have
\begin{align}
	\frac{\sum_{i=1}^{13}x_i^2+y_i^2}{(\sum_{i=1}^{13}x_i^2)(\sum_{i=1}^{5}y_i^2)-(\sum_{i=1}^{13}x_iy_i)^2}>\frac{2}{(2M^2+4(\frac{M}{2}-1))^2}. \label{30}
\end{align}
If $5<F_p<13$, all of the greater signs ($>$) in \eqref{29} and \eqref{30} will be replaced by greater or equal sign ($\ge$), with equality happening for any two subsets of $\mathcal{S}_{p_1}\bigcup\mathcal{S}_{p_2}$.

For $14\le F<21$ and other steps the proof is exactly same as $5<F_p\le13$. When all of antennas in the outer most rectangle are selected, then the proof of the next step that is the selection of the four antennas at the corner of the second most outer rectangle, is same as $F_p\le5$. Then the process of rest antennas in the same rectangle is same as $5<F_p\le13$, and so on. This process can be repeated for any desired $F_p$. 
\section{Proof of Theorem \ref{T2}} \label{App2}
As the main reference for the system, the antenna that is considered as the reference point has to be in the selected set. Then, the $CRLB_\theta$ for any set of three selected antennas $\{(0,0),(x_1,y_1),\\(x_2,y_2)\}$, will have the following form 
	\begin{align}
		CRLB_{\theta,3}=A\times\underbrace{\left(\frac{(x_1+\alpha y_1)^2+(x_2+\alpha y_2)^2}{(x_1y_2-x_2y_1)^2}\right)}_{Q_\theta(x_1,y_1,x_2,y_2)}, \label{p1}
	\end{align}	
	where $A=\frac{\cos^2(\theta)}{2\rho\beta^2|h_{d}|^2\sin^2(\varphi)}$ and it is independent from selected antennas. So, $Q_\theta(x_1,y_1,x_2,y_2)$ should be minimized with respect to (w.r.t) $x_1$, $y_1$, $x_2$ and $y_2$
	\begin{subequations}
	\begin{align}
		&\min_{x_1,y_1,x_2,y_2} Q_\theta(x_1,y_1,x_2,y_2), \label{p2}\\
		&s.t. \hspace{2mm} x_1,y_1,x_2,y_2 \in \{-\frac{M}{2},\ldots,0, \ldots, \frac{M}{2}\},\\
		&\hspace{6mm} (x_1,y_1)\neq(ax_2,ay_2)\hspace{3mm} for\hspace{3mm} a\in \mathbb{R} , \label{line}\\ %\setminus \{0\}
		&\hspace{6mm} (x_1,y_1)\neq(0,0), \hspace{3mm} (x_2,y_2)\neq(0,0) , \label{line2}
	\end{align}	
	\end{subequations}
	where \eqref{line} and \eqref{line2} ensure that an antenna is not selected twice, and three selected antennas do not compose a line.
	
	For $0<\theta\le\frac{\pi}{4}-\theta_0$, first we relax the integer constraint only on $x_2$ and prove that any other choice than $(x_1^*, y_1^*, x_2^*, y_2^*)=(-\frac{M}{2}, \frac{M}{2},\frac{M(1-2\alpha)}{2}, \frac{M}{2})$ results in higher $Q_\theta(x_1,y_1,x_2,y_2)$. Then the problem boils down to the choice between two upper and lower integers in the vicinity of $x_2$, that can be efficiently solved by directly checking which one results in lower value for $Q_\theta(x_1,y_1,x_2,y_2)$. 
	
	By replacing $(x_1^*, y_1^*, x_2^*, y_2^*)=(-\frac{M}{2}, \frac{M}{2},\frac{M(1-2\alpha)}{2}, \frac{M}{2})$ in $Q_\theta$, we have 
	\begin{align}
		Q_\theta(x_1^*, y_1^*, x_2^*, y_2^*)=\frac{2}{M^2}.
	\end{align}
	Now we show that for any combination of $(x_1, y_1, x_2, y_2)$,
	\begin{equation}
		\frac{(x_1+\alpha y_1)^2+(x_2+\alpha y_2)^2}{(x_1y_2-x_2y_1)^2}\ge\frac{2}{M^2}. \label{Main1}
	\end{equation}
	By multiplying both sides of \eqref{Main1} by $M^2(x_1y_2-x_2y_1)^2$, we have
	\begin{align}
		&M^2((x_1+\alpha y_1)^2+(x_2+\alpha y_2)^2)\nonumber\\
		&=M^2(x_1^2 +\alpha^2y_1^2+2\alpha x_1y_1+x_2^2+\alpha^2y_2^2+2\alpha x_2y_2)\nonumber\\
		&= 2\frac{M^2}{4}x_1^2 +M^2\alpha^2y_1^2+2M^2\alpha x_1y_1+2\frac{M^2}{4}x_2^2+M^2\alpha^2y_2^2+2M^2\alpha x_2y_2+2\frac{M^2}{4}(x_1^2+x_2^2)%\nonumber\\&+2((\frac{M^2}{4}-y_2^2)x_1^2+(\frac{M^2}{4}-y_1^2)x_2^2)
		\nonumber\\
		&= M^2\alpha^2y_1^2+M^2\alpha^2y_2^2+2M^2\alpha x_1y_1+2M^2\alpha x_2y_2+2x_2^2y_1^2+2x_1^2y_2^2+2\frac{M^2}{4}(x_1-x_2)^2\nonumber\\&-4x_1x_2y_1y_2+2((\frac{M^2}{4}-y_2^2)x_1^2+(\frac{M^2}{4}-y_1^2)x_2^2)+4 x_1x_2(\frac{M^2}{4}+y_1y_2). \label{Dif11}
	\end{align}
	Now we show the fact that the following phrase is always non-negative
	\begin{align}
		A(M,\alpha)&=M^2\alpha^2y_1^2+M^2\alpha^2y_2^2+2\alpha M^2 x_1y_1+2\alpha M^2 x_2y_2+2((\frac{M^2}{4}-y_2^2)x_1^2+(\frac{M^2}{4}-y_1^2)x_2^2)\nonumber\\&+2\frac{M^2}{4}(x_1-x_2)^2+4 x_1x_2(\frac{M^2}{4}+y_1y_2). 
	\end{align}
	By rewriting $A(M,\alpha)$ we have
	\begin{align}
		A(M,\alpha)=M^2(y_1^2+y_2^2)\alpha^2+2M^2(x_1y_1+x_2y_2)\alpha+(M^2(x_1^2+x_2^2)-2(x_1y_2-x_2y_1)^2). \label{alpha33}
	\end{align} 
	After some algebraic operations, the discriminant of $A(M,\alpha)$ will be obtained as
	\begin{align}
		\Delta_A&=4M^2(x_1y_2-x_2y_1)^2(2(y_1^2+y_2^2)-M^2)\le0, \label{DA}
	\end{align}
	where the last inequality results from the fact that $|y_1|\le\frac{M}{2}$ and $|y_2|\le\frac{M}{2}$. Because $\Delta_A\le0$ (and $M^2(y_1^2+y_2^2)>0$), for any value of $\alpha$ any combination of $x_1$, $x_2$, $y_1$ and $y_2$, $A(M,\alpha)$ will be non-negative. Therefore, we continue \eqref{Dif11} as
	\begin{align}
		M^2((x_1+\alpha y_1)^2+(x_2+\alpha y_2)^2)&= A(M,\alpha)+2x_2^2y_1^2+2x_1^2y_2^2-4x_1x_2y_1y_2\nonumber\\
		&\ge2x_2^2y_1^2+2x_1^2y_2^2-4x_1x_2y_1y_2=2(x_1y_2-x_2y_1)^2.		 \label{easy2}
	\end{align}
	So, $(x_1^*, y_1^*, x_2^*, y_2^*)=(-\frac{M}{2}, \frac{M}{2},\frac{M(1-2\alpha)}{2}, \frac{M}{2})$ is the optimal choice to minimize $Q_\theta(x_1,y_1,x_2,y_2)$. However, $\frac{M(1-2\alpha)}{2}$ may not be an integer value. In order to decide whether higher or lower integer is the better choice, we simply check both of them in the function of $Q_\theta(x_1,y_1,x_2,y_2)$. The $Q_\theta(-\frac{M}{2}, \frac{M}{2}, x, \frac{M}{2})$ will be 
	\begin{align}
		Q_\theta(-\frac{M}{2}, \frac{M}{2}, x, \frac{M}{2})=\frac{(1-\alpha)^2+(\frac{2}{M}x+\alpha)^2}{(\frac{M}{2}+x)^2}. \nonumber
	\end{align}
	
	As $\theta\to\frac{\pi}{4}$, $\frac{M(1-2\alpha)}{2}\to-\frac{M}{2}$, but because one antenna cannot be selected twice, the system will have to choose $1-\frac{M}{2}$ for $x_2$. The selection of $(-\frac{M}{2}, \frac{M}{2},1-\frac{M}{2}, \frac{M}{2})$ happens when $\theta\ge \tan^{-1}(\frac{M-1}{M})$. Although this choice is optimal for a certain range of $\theta$, after a point it will not be the best choice and $(x_1, y_1, x_2, y_2)=(-\frac{M}{2}, \frac{M}{2}-1,1-\frac{M}{2}, \frac{M}{2})$ results in lower $Q_\theta(x_1,y_1,x_2,y_2)$. To obtain the range of $\theta$ that the latter set is the optimal choice, we have to solve the following inequality for $\alpha$
	\begin{align}
		Q_\theta(-\frac{M}{2}, \frac{M}{2}-1,1-\frac{M}{2}, \frac{M}{2})< Q_\theta(-\frac{M}{2}, \frac{M}{2},1-\frac{M}{2}, \frac{M}{2}).
	\end{align}
	By replacing the actual values of both sides, we have
	\begin{align}
		\frac{(\alpha-1)^2(\frac{M^2}{2}-M+1)+2\alpha}{(M-1)^2}<(\alpha-1)^2+\frac{(M(\alpha-1)+2)^2}{M^2}. \label{ineq}
	\end{align}
	After some algebraic operations, \eqref{ineq} simplifies to
	\begin{align}
		B(M,\alpha)&=M^2(3M^2-6M+2)\alpha^2-2M(3M^3-10M^2+12M-4)\alpha\nonumber\\&+(M-2)(3M-2)(M^2-2M+2)>0. \label{ineq2}
	\end{align}
	Calling $\alpha_{b_1}$ the smaller root and $\alpha_{b_2}$ the bigger root of $B(M,\alpha)$, for $M\ge1+\frac{1}{\sqrt{3}}$ \footnote{The roots of $(3M^2-6M+2)$ are $1\pm\frac{1}{\sqrt{3}}$ and for $M\ge1+\frac{1}{\sqrt{3}}$ , the multiplier of $\alpha^2$ will be always positive.} (which is the case because $M\ge2$), the inequality of \eqref{ineq2} is held when either $\alpha\le\alpha_{b_1}$ or $\alpha_{b_2}\le\alpha$. To find out which root indicates the value of $\theta_0$, after which $(x_1, y_1, x_2, y_2)=(-\frac{M}{2}, \frac{M}{2}-1,1-\frac{M}{2}, \frac{M}{2})$ is the optimal choice, we investigate the sign of $B(M,\alpha)$ in a critical point in which $x_2=\frac{M(1-2\alpha)}{2}$ is an integer and therefore $(x_1, y_1, x_2, y_2)=(-\frac{M}{2}, \frac{M}{2},\frac{M(1-2\alpha)}{2}, \frac{M}{2})$ is still the optimal choice. This point is $\alpha=\frac{M-1}{M}$. By replacing $\alpha=\frac{M-1}{M}$ in $B(M,\alpha)$ and simplifying it, we obtain
	\begin{align}
		B(M,\frac{M-1}{M})=-M^2-2M+2. \label{sign1}
	\end{align}
	The roots of $B(M,\frac{M-1}{M})$ are $-\sqrt{3}-1$ and $\sqrt{3}-1$. So, for $M\ge2$, \eqref{sign1} will always be negative (sign of $M^2$). This means that $\alpha=\frac{M-1}{M}$ is between the roots of $B(M,\alpha)$ (this is because for $M\ge2$ the sign of $B(M,\alpha)$ is negative only between its root). We know that for $\alpha\le\frac{M-1}{M}$, the optimal choice is $(x_1, y_1, x_2, y_2)=(-\frac{M}{2}, \frac{M}{2},\frac{M(1-2\alpha)}{2}, \frac{M}{2})$ and therefore, the minimum $\alpha$ that changes the optimal choice, must be bigger than $\frac{M-1}{M}$. So, the minimum $\alpha$ after which the optimal choice becomes $(x_1, y_1, x_2, y_2)=(-\frac{M}{2}, \frac{M}{2}-1,1-\frac{M}{2}, \frac{M}{2})$ is the bigger root of $B(M,\alpha)$ (solved for $\alpha$). By noting that $\alpha=\tan(\theta)$, we obtain $\theta_0=\tan^{-1}(\alpha_{b_2})$.
	
	For $\theta_1\le\theta<\frac{\pi}{2}$, we relax the integer constraint only on $y_2$ to show that $(x_1, y_1, x_2, y_2)=(-\frac{M}{2}, \frac{M}{2},-\frac{M}{2}, \frac{M(2-\alpha)}{2\alpha})$ is the optimal choice. By replacing $(x_1^*, y_1^*, x_2^*, y_2^*)=(-\frac{M}{2}, \frac{M}{2},-\frac{M}{2}, \frac{M(2-\alpha)}{2\alpha})$ in \eqref{p1} we have
	\begin{align}
		f(x_1^*, y_1^*, x_2^*, y_2^*)=\frac{2\alpha}{M^2}. \label{3rdcrlb}
	\end{align}
	The process to show that any other choice than $(x_1, y_1, x_2, y_2)=(-\frac{M}{2}, \frac{M}{2},-\frac{M}{2}, \frac{M(2-\alpha)}{2\alpha})$ results in higher $Q_\theta$ is same as the process of \eqref{Dif11}. The only difference is that instead of showing $A(M,\alpha)\ge0$, we should show that the following equation is non-negative:
	\begin{align}
		C(M,\alpha)=(M^2(y_1^2+y_2^2)-2(x_1y_2-x_2y_1)^2)\alpha^2+2M^2(x_1y_1+x_2y_2)\alpha+M^2(x_1^2+x_2^2). \label{Bma}
	\end{align}
	The discriminant of $C(M,\alpha)$ will be obtained as
	\begin{align}
		\Delta_C&=4M^2(x_1y_2-x_2y_1)^2(2(x_1^2+x_2^2)-M^2)\le0. \label{DB}
	\end{align}
	Same as \eqref{DA}, $\Delta_C$ in \eqref{DB} is always non-positive, confirming that $C(M,\alpha)\ge0$.
	
	When $\theta$ is around $\frac{\pi}{4}$, $\frac{M(2-\alpha)}{2\alpha}$ is close to $\frac{M}{2}$ but as one antenna cannot be selected twice, system has to select $\frac{M}{2}-1$ for $y_2$. This obligation causes the choice of $(1-\frac{M}{2}, \frac{M}{2},-\frac{M}{2}, \frac{M}{2}-1)$ to be the optimal choice for a certain range (same as the case of $\theta_0<\theta<\frac{\pi}{4}$). To obtain this range, we have to solve the following inequality
	\begin{align}
		Q_\theta(-\frac{M}{2}, \frac{M}{2}-1,1-\frac{M}{2}, \frac{M}{2})< Q_\theta(-\frac{M}{2}, \frac{M}{2},-\frac{M}{2}, \frac{M}{2}-1).
	\end{align}
	By replacing the actual values of both sides and simplifying the inequality we obtain
	\begin{align}
		D(M,\alpha)&=(M-2)(3M-2)(M^2-2M+2)\alpha^2-2M(3M^3-10M^2+12M-4)\alpha\nonumber\\&+M^2(3M^2-6M+2)>0. \label{ineq3}
	\end{align}
	Calling $\alpha_{d_1}$ the small root and $\alpha_{d_2}$ the big root of $D(M,\alpha)$, for $M>2$ the inequality of \eqref{ineq3} is held when either $\alpha\le\alpha_{d_1}$ or $\alpha_{d_2}\le\alpha$. To find out which root indicates the value of $\theta_1$, before which $(x_1, y_1, x_2, y_2)=(-\frac{M}{2}, \frac{M}{2}-1,1-\frac{M}{2}, \frac{M}{2})$ is the optimal choice, we investigate the sign of $D(M,\alpha)$ in a critical point in which $y_2=\frac{M(2-\alpha)}{2\alpha}$ is an integer and therefore $(x_1, y_1, x_2, y_2)=(-\frac{M}{2}, \frac{M}{2},-\frac{M}{2}, \frac{M(2-\alpha)}{2\alpha})$ is still the optimal choice. This point is $\alpha=\frac{M}{M-1}$. By replacing $\alpha=\frac{M}{M-1}$ in $D(M,\alpha)$ and simplifying it, we obtain
	\begin{align}
		D(M,\frac{M}{M-1})=(-M^2-2M+2)(\frac{M}{M-1})^2. \label{sign2}
	\end{align}
	As $(\frac{M}{M-1})^2$ is always positive and the roots of $(-M^2-2M+2)$ are $-1\pm\sqrt{3}$, for $M>1$, $D(M,\frac{M}{M-1})<0$. This means that $\alpha=\frac{M}{M-1}$ is between the roots of \eqref{ineq3}. We know that for $\alpha\ge\frac{M}{M-1}$, the optimal choice is $(x_1, y_1, x_2, y_2)=(-\frac{M}{2}, \frac{M}{2},-\frac{M}{2}, \frac{M(2-\alpha)}{2\alpha})$ and therefore, the maximum $\alpha$ before which the optimal choice is $(x_1, y_1, x_2, y_2)=(-\frac{M}{2}, \frac{M}{2}-1,1-\frac{M}{2}, \frac{M}{2})$ has to be the smaller root of \eqref{ineq3}. By noting that $\alpha=\tan(\theta)$, we obtain $\theta_1=\tan^{-1}(\alpha_{d_1})$.
	
	By comparing the \eqref{ineq2} and \eqref{ineq3} we see that both have same coefficient for $\alpha$, and the $\alpha^2$ coefficient in \eqref{ineq2} is the constant term of \eqref{ineq3} and vice versa. We know that the roots of $ax^2+bx+c$ are $x_{1,2}=\frac{-b\pm\sqrt{b^2-4ac}}{2a}$ and the roots of $cx^2+bx+a$ are $x_{3,4}=\frac{-b\pm\sqrt{b^2-4ac}}{2c}=\frac{a}{c}x_{1,2}$. Therefore, $\alpha_{d_1}=\frac{\alpha_{b_1}M^2(3M^2-6M+2)}{(M-2)(3M-2)(M^2-2M+2)}$ and  $\theta_1=\tan^{-1}(\alpha_{d_1})=\tan^{-1}(\frac{M^2(3M^2-6M+2)\alpha_{b_1}}{(M-2)(3M-2)(M^2-2M+2)})$.
\bibliography{References}
\bibliographystyle{ieeetr}
\end{document}